\documentclass[aps,prd,reprint,twocolumn,showpacs,superscriptaddress,
longbibliography,nofootinbib]{revtex4-2}  
\usepackage{array}[=2016-10-06]
\usepackage{tabularx}
\makeatletter
\def\hlinewd#1{%
\noalign{\ifnum0=`}\fi\hrule \@height #1 %
\futurelet\reserved@a\@xhline}
\makeatother
\usepackage{scrextend}
\usepackage{amsmath}
\usepackage{amssymb}
\usepackage{epsfig}
\usepackage{graphicx}
\usepackage{hyperref}
\usepackage{dcolumn}
\usepackage{slashed}
\usepackage{color}
\usepackage{rotating}
\usepackage[margin=0.9in,a4paper]{geometry}
\usepackage[table,xcdraw,dvipsnames]{xcolor}
\usepackage[utf8]{inputenc}
\usepackage{colortbl}
\usepackage[normalem]{ulem}
\usepackage{mathrsfs} 

\usepackage[vcentermath]{youngtab}
\usepackage{ytableau}

\usepackage{calligra} 

\definecolor{nicered}{rgb}{0.7,0.1,0.1}
\definecolor{nicegreen}{rgb}{0.1,0.5,0.1}
\definecolor{red}{rgb}{1.0, 0, 0}

\newcommand{\N}{{\cal N}}
\newcommand{\Y}{{\cal Y}}
\newcommand{\A}{{\cal A}}

\newcommand{\SU}{{\rm SU}}

\newcommand{\U}{{\rm U}}
\newcommand{\F}{{\cal F}}

\renewcommand{\[}{\left[}

\renewcommand{\(}{\left(}
\renewcommand{\)}{\right)}

\newcommand{\bdm}{\begin{displaymath}}
\newcommand{\edm}{\end{displaymath}}
\newcommand{\bea}{\begin{eqnarray}}
\newcommand{\eea}{\end{eqnarray}}
\newcommand{\Q}{\mathcal{Q}}
\newcommand{\mQ}{\mathcal{Q}}
\newcommand{\mB}{\mathcal{B}}
\newcommand{\mN}{\mathcal{N}}
\newcommand{\mM}{\mathcal{M}}

\definecolor{darkgreen}{rgb}{0.1,0.7,0.1}
\definecolor{nicered}{rgb}{0.7,0.1,0.1}
\definecolor{nicegreen}{rgb}{0.1,0.5,0.1}
\definecolor{red}{rgb}{1.0, 0, 0}
\definecolor{niceblue}{rgb}{0,0,0.8}
\definecolor{red}{rgb}{1.0, 0, 0}
\hypersetup{
 colorlinks,
 citecolor=nicegreen,
 linkcolor=nicered,
 urlcolor=nicered,
 pdftitle={Reconciling axion quality with post-inflation cosmology},
 pdfauthor={L. Di Luzio, S. Di Valeriano, M. Nardecchia, E. Nardi}
}

\def\eq#1{{Eq.~(\ref{#1})}}

\def\fig#1{{Fig.~\ref{#1}}}

\def\Table#1{{Table~\ref{#1}}}

\def\app#1{{App.~\ref{#1}}}

\def\Tr{\mbox{Tr}\,}

\def\det{\mbox{det}\,}
\def\arg{\mbox{arg}\,}

\def\gsim{\raise0.3ex\hbox{$\;>$\kern-0.75em\raise-1.1ex\hbox{$\sim\;$}}}
\def\lsim{\raise0.3ex\hbox{$\;<$\kern-0.75em\raise-1.1ex\hbox{$\sim\;$}}}

\def\mb[#1]{\mathbf{#1}}

\definecolor{LightCyan}{rgb}{0.88,1,1}
\definecolor{piggypink}{rgb}{0.99, 0.87, 0.9}
\definecolor{applegreen}{rgb}{0.55, 0.71, 0.0}
\definecolor{darkpastelgreen}{rgb}{0.01, 0.75, 0.24}
\definecolor{green-yellow}{rgb}{0.68, 1.0, 0.18}

\newcommand{\beq}{\begin{equation}}
\newcommand{\eeq}{\end{equation}}
\newcommand{\beqa}{\begin{eqnarray}}
\newcommand{\eeqa}{\end{eqnarray}}

\newcommand{\GeV}{{\, \rm GeV}}

\newcommand{\eqn}[1]{Eq.~(\ref{#1})}

\definecolor{revisionblue}{RGB}{0,76,153}
\definecolor{todoorange}{RGB}{190,85,0}
\definecolor{thresholdviolet}{RGB}{135,30,135}
\definecolor{collabcrimson}{RGB}{170,0,70}
\newif\ifshownotes
\shownotesfalse

\begin{document}

\title{Reconciling axion quality with post-inflation cosmology}

\author{Luca Di Luzio}
\email{luca.diluzio@pd.infn.it}
\affiliation{\small \it 
INFN Sezione di Padova, Via Francesco Marzolo 8, 35131 Padova, Italy}
\author{Samuele Di Valeriano}
\email{sdivaler@sissa.it}
\affiliation{\small \it SISSA, International School for Advanced Studies, Via Bonomea 265, I-34136 Trieste, Italy}
\author{Marco Nardecchia}
\email{marco.nardecchia@roma1.infn.it}
\affiliation{\small \it Physics Department and INFN Sezione di Roma La Sapienza, Piazzale Aldo Moro 5, 00185 Roma, Italy}
\author{Enrico Nardi}
\email{enrico.nardi@lnf.infn.it}
\affiliation{\small \it Istituto Nazionale di Fisica Nucleare, Laboratori Nazionali di Frascati, C.P.~13, 00044 Frascati,
Italy}
\affiliation{\small \it Laboratory of High Energy and Computational Physics, NICPB, R\"avala 10, 10143, Tallinn, Estonia}

\begin{abstract}
We address the 
axion quality/cosmology tension that often plagues QCD axion models in post-inflationary  
Peccei-Quinn (PQ) breaking scenarios.  
In the proposed  framework, the PQ symmetry
emerges accidentally from an $\SU(\mathcal{N})_L\times\SU(\mathcal{N})_R$ gauge symmetry spontaneously broken to $\SU(\mathcal{N})_{L+R}$, and its quality is protected because PQ breaking first appears at operator dimension $\mathcal{N}$.
 Moreover, the  construction avoids the domain-wall
problem, ensures that
no stable fractionally charged relic survives $\SU(\mathcal{N})_{L+R}$ confinement,  
and  remains free of
Landau poles below the scale of PQ-breaking operators.
The exotic quarks required to generate  the PQ anomaly hadronize into 
unstable mesons and  cosmologically stable neutral baryons. 
However, plausible arguments indicating a strong suppression of baryon formation 
in $\SU(\mathcal{N})$ gauge theories at large $\mathcal{N}$, 
suggest that their contribution 
 to the dark matter energy density remains subdominant.  
Combining cosmological constraints with PQ quality and
perturbativity requirements yields a viable parameter space in which the
axion makes up most of the dark matter and its mass remains predictable.
\end{abstract}

\maketitle

\section{Introduction}
\label{sec:intro}
The Peccei--Quinn (PQ) mechanism provides the most compelling dynamical solution to the
strong CP problem~\cite{Peccei:1977hh,Peccei:1977ur}.  Its associated
Nambu--Goldstone boson, the axion~\cite{Wilczek:1977pj,Weinberg:1977ma}, is
also a well-motivated dark matter (DM) candidate~\cite{Preskill:1982cy,Abbott:1982af,
Dine:1982ah}.  The experimental program to search for the axion has advanced rapidly and is beginning
to probe theoretically motivated parameter space; see
Refs.~\cite{Irastorza:2018dyq,DiLuzio:2020wdo,Sikivie:2020zpn,
Adams:2022pbo} for reviews.

The construction of a complete axion model nevertheless faces a basic
ultraviolet (UV)
question.  In benchmark models the global $\U(1)_{\rm PQ}$ symmetry is imposed
by hand~\cite{Kim:1979if,Shifman:1979if,Zhitnitsky:1980tq,Dine:1981rt}. 
A fundamental theory should instead explain its origin as an accidental symmetry 
while ensuring that any source of explicit PQ symmetry breaking is suppressed enough 
to preserve the strong CP solution.
Specifically, the minimum of the axion potential must not shift from the CP-conserving minimum by more than
$\langle a\rangle/f_a \lesssim 10^{-10}$, where $f_a$ denotes the axion decay constant.  
Consequently, the PQ mechanism demands an exceptionally high-quality accidental global symmetry.
This is the axion quality problem~\cite{Georgi:1981pu,Dine:1986bg,
Barr:1992qq,Kamionkowski:1992mf,Holman:1992us,Ghigna:1992iv}, which has
motivated a broad range of theoretical constructions;
see, e.g., Refs.~\cite{Randall:1992ut,Dobrescu:1996jp,Babu:2002ic,
Redi:2016esr,Fukuda:2017ylt,DiLuzio:2017tjx,Bonnefoy:2018ibr,
Lillard:2018fdt,Gavela:2018paw,Lee:2018yak,Ardu:2020qmo,Yin:2020dfn,
DiLuzio:2020qio,Contino:2021ayn,DiLuzio:2025jhv}.

Building a viable high-quality axion model is especially restrictive when the PQ symmetry breaks, or is temporarily restored, after inflation. 
We refer to this scenario as post-inflationary PQ breaking.
In this scenario, axion strings formed at the PQ phase transition (PT) become interconnected by domain walls during the QCD confinement 
epoch. In most cases, this string-wall network is stable and quickly dominates the cosmological energy density, which is 
phenomenologically unacceptable. Furthermore, UV completions of the PQ mechanism that address the quality problem 
often imply the existence of new cosmologically stable particles. Some of these relics can carry an electric charge, and in particular  
in the case of fractionally charged states, experimental searches constrain their abundances to values many orders of magnitude 
below that of ordinary matter~\cite{Perl:2009zz,Perl:2004qc,Halyo:1999wq}.  
In pre-inflationary scenarios  these problems  are absent since strings, domain walls, and heavy dangerous relics can all be inflated away. 
 However,  this option sacrifices the predictive relationship between the axion mass and the observed DM abundance,
 since the latter depends on the initial value of the axion field. 
In post-inflationary scenarios, avoiding the domain-wall problem requires specific constructions where  
each axion string is attached to a single wall~\cite{Sikivie:1982qv}. In   
this case the string-wall network is unstable and quickly decays. Ensuring the absence of dangerous 
heavy relics also demands careful model-building.  
However, a compelling feature  of post-inflationary scenarios lies in their enhanced predictability, since they can potentially 
yield a precise relationship between the axion mass and the observed DM abundance.   
  Currently, the primary uncertainty in establishing this quantitative relation stems from the string and wall contributions to 
  the axion relic density. However, recent numerical simulations have steadily advanced, 
  pushing 
  toward a definitive prediction~\cite{Hiramatsu:2012gg,Klaer:2017ond,Gorghetto:2018myk,
Gorghetto:2020qws,Buschmann:2021sdq,Saikawa:2024bta,Benabou:2024msj}.

As recently highlighted in Ref.~\cite{Lu:2023ayc}, the axion
quality/cosmology tension arises from demanding that a satisfactory  UV construction
protects the PQ quality to the required level, produces the minimal axion string needed to ensure  
that the string-wall network is unstable,   and does not yield dangerous cosmologically stable relics.
Known models that raise the dimension of PQ-violating operators often introduce large Standard Model (SM) multiplets.  
However, the augmented matter content can drive a SM gauge coupling to a Landau pole below the
scale $\Lambda_{\rm PQ}$  at which PQ violation is assumed to originate.  A satisfactory 
construction must therefore solve the PQ-quality, domain-wall, and dangerous relic problems, without
losing perturbative control below $\Lambda_{\rm PQ}$.\footnote{Here we focus on four-dimensional
field-theory setups. 
For extra-dimensional constructions addressing axion quality,
see e.g.~Refs.~\cite{Choi:2003wr,Reece:2025thc,Craig:2024dnl,Petrossian-Byrne:2025mto,Choi:2026kxu,FernandezNavarro:2026bed}.}

We build on the $\SU(\mN)_L\times\SU(\mN)_R$ model introduced in
Ref.~\cite{DiLuzio:2017tjx}.  In that construction, a  scalar $Y$ 
transforming in the bifundamental representation of dimension 
$\mN \times \mN$  breaks the gauge symmetry to the diagonal subgroup
 $\SU(\mN)_{L+R}$. The scalar potential $V(Y)$ carries  an accidental
PQ symmetry. The  lowest dimension operator that violates this symmetry  is $\det  Y$, that has  
dimension $\mN$,  implying that $\U(1)_{\rm PQ}$ becomes exact in the large $\mN$ limit.
The symmetry-breaking pattern $\SU(\mN)_L\times\SU(\mN)_R \times \U(1)_{\rm PQ} \to \SU(\mN)_{L+R}$ 
realizes the Lazarides--Shafi
gauge identification of axion vacua and produces the minimal axion string
required for an unstable string--wall network
~\cite{Lazarides:1982tw,Lu:2023ayc}.
We retain the fermion content of
Ref.~\cite{DiLuzio:2017tjx}, consisting of a QCD-colored fermion $\Q$ and three
QCD-singlet fermions $\Psi^a$ ($a=1,2,3$).  In that work, all these fermions
were assigned zero hypercharge, leading to stable fractionally charged QCD
hadrons.  For this reason, in order to inflate away 
all the dangerous relics,  in Ref.~\cite{DiLuzio:2017tjx} 
a pre-inflationary PQ breaking scenario was assumed.

A new ingredient of the present work is the hypercharge assignment of the heavy fermions, which allows for  
decay channels. 
Ref.~\cite{Lu:2023ayc} proposed a SM-like hypercharge assignment for
three replicated flavors of the QCD-colored $\Q$ field.  This neatly allows
all exotic relics to decay, but the proliferation of SM-charged matter
produces low-scale Landau poles in both the $\SU(3)_C$ and $\U(1)_Y$
couplings. 
In the present construction instead, we assign a hypercharge to the $\Psi^a$ fields,
while  the $\Q$'s  remain neutral.   Assigning  QCD-color and hypercharge to different
fermion  species reduces their multiplicity, limiting their  impact on the running of the SM gauge couplings while still allowing
for certain dimension-six   
operators that are needed to ensure the viability of the model. The presence of these operators 
leaves a single residual $\U(1)_{\Q-\Psi}$ symmetry, allowing for the decays of the heavier species into 
the lightest one.  The quality of this symmetry  is tied to that of $\U(1)_{\rm PQ}$, and 
therefore  the lightest baryon carrying this approximately conserved charge is cosmologically stable. 
 For $\mN \equiv 0\pmod 3$, this baryon can only have  integer
electric charge, and the mass spectrum of the new fermions can be arranged in 
such a way that the electric charge is zero.
At the scale  $\Lambda_{L+R}$  the diagonal $\SU(\mN)_{L+R}$ group confines, and 
the lightest $\mN$-quarks  surviving in the plasma hadronize in 
$\SU(\mN)_{L+R}$  singlets. In particular,  baryons can form as bound states 
of $\mN$ quarks. 
The nature of these baryons is determined 
by the type of quarks present in the plasma at the confining epoch.
 We consider  two orderings for the $\Psi$ and $\Q$ masses: 
$m_\Q<m_\Psi$ and $m_\Psi<m_\Q$.  In the first case, the lightest baryon $\mB_\Q\sim [\mQ^\mN]$ 
 is the electrically neutral,  and  carries QCD-color in a configuration of zero-triality.  
 At the QCD PT  it will get dressed by gluons into a color-singlet neutral baryon
 $\mB_0^{(g)}$.  
 In the second case, assuming  that the  
mass splitting between different constituent quark flavors is smaller than  the hypercharge self-energy 
contribution to the baryon mass,  an electrically neutral, QCD-singlet baryon $\mB_0$
is selected as the lightest, cosmologically  stable,  state. While this  relic avoids the strongest limits  
on   charged  stable particles, it is crucial to ensure that  its contribution to the cosmic energy density 
will not exceed that of DM. 
Baryon formation in $\SU(\mN)$ at large $\mN$ has recently been studied
 in Refs.~\cite{casimirletter,inprogress}, to which we refer for further details.   
 There it is suggested that  at  large $\mN$ a Casimir bottleneck 
strongly quenches baryon assembly.
This phenomenon could efficiently reduce the cosmological abundance of the stable $\mB_0$ baryons for 
 both mass orderings. 
Although the arguments presented in Refs.~\cite{casimirletter,inprogress} are not derived from first principles, the resulting framework incorporates the relevant physical ingredients in a well-motivated and reasonable way, allowing us to establish quantitative benchmarks.
 We require that $\mB_0$ relics will 
contribute at most at the level of ten percent  to the  DM energy density, so that the axion
remains the leading component, and the model maintains  its   predictivity  for the axion mass.
We  confront both mass orderings with the
PQ-quality and two-loop Landau-pole constraints, requiring that they are jointly respected.  
These requirements select $\mN=15$, $18$, and
$21$ as representative successive multiples of three 
for which all the constraints are satisfied. 
For $\mN\leq12$, both the  suppression of PQ violation
from the determinant operator and the suppression of  the stable-baryon yield from the  Casimir bottleneck  are 
not sufficient to satisfy the constraints. 
The $\mN=21$ case already permits
order-one Yukawa couplings and a common PQ-breaking and flavor-transfer
scale.  Increasing $\mN$ further brings no qualitatively new phenomenological
advantage for the purposes of this work, while continuing to worsen the SM
running.

The paper is organized as follows.  In Sec.~\ref{sec:axion-quality} we review
the  axion sector of the $\SU(\mN)_L\times\SU(\mN)_R$ model. In Sec.~\ref{sec:accidental-symmetries}
we discuss its accidental symmetries.  In Sec.~\ref{sec:spectrum-bound-states}
we describe  the spectrum and identify the stable baryon candidates.
Sec.~\ref{sec:postinfl-cosmology} analyzes the post-inflationary
cosmology of the model for the two mass orderings.
In Sec.~\ref{sec:uv-constraints} we combine the
cosmological requirements with the PQ-quality and Landau-pole constraints to identify the 
values of  $\mN$ that yield viable models, and
we conclude in Sec.~\ref{sec:concl}. The allowed hypercharges for the $\Psi$ states are listed in
App.~\ref{sec:hypercharge}. The pseudo-Nambu--Goldstone bosons (pNGBs) and the
$\eta'$ sector are analyzed in App.~\ref{app:pngb}.  
In App.~\ref{app:confinement-hadronization} we summarize the main 
results from  Refs.~\cite{casimirletter,inprogress} regarding   
the   network of recombination reactions  responsible for hadronization 
and the Casimir bottleneck. 
The two-loop renormalization group (RG)
equations are given in App.~\ref{sec:two-loop-running}.

\section{Axion quality from a $G_{LR}^{(\N)}$ bifundamental}
\label{sec:axion-quality}

We begin by recalling the main ingredients of the model in Ref.~\cite{DiLuzio:2017tjx}. 
We consider the gauge group $G_{LR}^{(\N)} \equiv \SU(\N)_{L} \times \SU(\N)_{R}$, with a 
complex scalar $Y$ that transforms as a bifundamental $(\N, \bar \N)$ of $G_{LR}^{(\N)}$. 
For $\N > 4$ the renormalizable scalar potential, $V(Y)$,  contains only invariants of the type 
$\Tr [YY^\dag]$ and $\Tr [YY^\dag YY^\dag]$, thus featuring an accidental global symmetry, 
$Y \to e^{i\alpha/\N} Y$. 
We denote this symmetry as $\U(1)_{\rm PQ}$, anticipating its later identification with a PQ symmetry anomalous under QCD.
The matrix $Y^c$ of constant
background values of $Y(x)$ can be written, 
via a rigid $G_{LR}^{(\N)} $ transformation, 
as
$Y^c = (v_a/\sqrt{2})
\hat \Phi\hat Y$,  
where $\hat Y$ 
is diagonal with real
non-negative entries, normalized such that $\Tr \hat Y^2 =1$, and 
$\hat\Phi$ 
is a diagonal matrix of phases 
such that 
$\log\det(\hat\Phi)=i\,\arg\det(Y^c)$,  
with $\arg\det(Y^c)\in[0,2\pi)$ being an angular variable.

For an appropriate choice of the potential parameters \cite{DiLuzio:2017tjx}
it is possible to obtain 
the vacuum expectation value 
(VEV) 
configuration $\hat Y = \frac{1}{\sqrt N}\mathrm{diag}(1,\ldots, 1)$, 
so that the little group is given by the diagonal subgroup $\SU(\N)_{L+R}$.
The breaking pattern $G_{LR}^{(\N)} \times \U(1)_{\rm PQ} \to \SU(\N)_{L+R}$, 
is such that also the accidental global $\U(1)_{\rm PQ}$ is spontaneously broken, leading to a Goldstone boson 
(to be identified later on with the QCD axion).\footnote{The alternative breaking pattern 
$G_{LR}^{(\N)}  \to G_{LR}^{(\N-1)}$ 
does not yield an elementary axion but instead gives rise to a composite one, as can be shown through an anomaly-matching argument~\cite{DiLuzio:2017tjx}. 
In the following, we restrict our attention to the elementary axion solution.}  
The canonically
normalized axion field is \cite{DiLuzio:2017tjx}
\begin{equation}
 \label{eq:axfielddef}
a(x)\equiv f_a\,\arg\det Y(x),
\qquad f_a\equiv\frac{v_a}{\N}\,.
\end{equation}
Since $\arg\det Y$ is $2\pi$ periodic, the axion is identified
under $a\to a+2\pi f_a$.

The leading operator that explicitly breaks the accidental $\U(1)_{\rm PQ}$ symmetry appears at dimension-$\N$ and is given by the determinant, 
$\mathcal{D}\equiv\det Y$.
Clearly, bi-unitary transformations acting on $Y$ leave $\mathcal D$ invariant, while under a $\U(1)_{\rm PQ}$ transformation, $Y \to e^{i\alpha/N} Y$, we get $\mathcal{D}\rightarrow e^{i\alpha}\mathcal{D}$. 
Therefore, the lowest-dimensional operator that breaks the $\U(1)_{\rm PQ}$ symmetry takes the form
\begin{equation}
\label{VD}
  \!  V_D = \frac{k_D}{\Lambda_{\rm PQ}^{\,N-4}}\mathcal{D} + \mathrm{h.c.}
        = \frac{2|k_D|}{\Lambda_{\rm PQ}^{\,N-4}}\,D
          \cos\!\left[\varphi + \frac{a(x)}{f_a}\right] \, ,
\end{equation}
Here $D\equiv|\mathcal D|$, $a(x)/f_a=\arg\mathcal D$, and
$\varphi\equiv\arg(k_D)$.
$\Lambda_{\rm PQ}$ denotes the scale associated with explicit breaking of the $\U(1)_{\rm PQ}$ symmetry. 
In the following, we assume $v_a \ll \Lambda_{\rm PQ} \leq M_{\rm Pl} = 1.2 \times 10^{19}$ GeV.

The minimum of $V_D$ is obtained for
$\langle a\rangle/f_a=\pi-\varphi$,
thereby jeopardizing the PQ solution to the strong CP problem,   
and the corresponding shift in the ground-state energy is
\begin{equation}
  \label{eq:DeltaV}
\Delta V = v_a^4 \frac{2|k_D|}{(2N)^{\N/2}}
\left(\frac{v_a}{\Lambda_{\rm PQ}}\right)^{\N-4} \, .   
\end{equation}
Thus, in the breaking  $G_{LR}^{(\N)} \times \U(1)_{\rm PQ} \to \SU(\N)_{L+R}$,
$\N^2-1$ of the initial $2(\N^2-1) + 1$ generators are left unbroken,
$\N^2 -1$ are spontaneously broken with the corresponding Goldstone modes eaten by
gauge bosons that acquire masses of $\mathcal{O}(v_a)$, while the Goldstone boson of the global
$\U(1)_{\rm PQ}$ acquires a tiny mass of $\mathcal{O}(\sqrt{\Delta V}/v_a)$ because of the
explicit breaking in \eq{VD}. 
The suppression of this contribution at large $\N$ is the key to preserving the quality of the $\U(1)_{\rm PQ}$ symmetry.

\subsection{Fermionic sector and strong CP} 

A solution to the strong CP problem can be implemented by introducing
fermions carrying color. 
Following \cite{DiLuzio:2017tjx}, we introduce four fermion multiplets
transforming under $G_{LR}^{(\N)}$ as: $\Q_L \sim (\N,1)$, $\Q_R \sim
(1,\N)$, $\Psi_L \sim (\bar \N,1)$ $\Psi_R \sim (1,\bar \N)$.  Since they
can be combined into real representations of $\SU(\N)_{L,R}$, 
there are no gauge anomalies.  Gauge
symmetry allows for Yukawa couplings of the form $\bar \Q_L Y \Q_R +
\bar\Psi_L Y^\dagger \Psi_R$. They preserve the $Y$
rephasing symmetry if the fermions are transformed chirally with
$\U(1)_{\rm PQ}$ charges satisfying: $\chi_{\Q_L} - \chi_{\Q_R} = - (\chi_{\Psi_L} -\chi_{\Psi_R})$. 
The opposite sign of the two
charge differences (a consequence of the requirement of gauge
anomalies cancellation) ensures the absence of $\U(1)_{\rm PQ}$-$\SU(\N)_{L,R}$
mixed anomalies.\footnote{This point is crucial for the PQ solution, as it guarantees that the axion VEV cancels $\theta_{\rm QCD}$ rather than the $\SU(\N)_{L,R}$ $\theta$-terms.} 
We  assign $\Q_{L,R}$ to the fundamental representation of
QCD-color $\SU(3)_C$.  The contribution of their  three color components 
to the $\SU(\N)_{L,R}$  gauge anomalies is canceled by three color-singlet 
$\Psi_{L,R}^a$ with flavor $a=1,2,3$.  Since these states do not carry color, there is no compensating cancellation of
the $\Q_{L,R}$ contribution to the global $\U(1)_{\rm PQ}$-$\SU(3)_C$ anomaly, so that  
$\U(1)_{\rm PQ}$ has all the features required to play the role of  a PQ symmetry.

The field content of the model is summarized in \Table{tab:NewMatterContent}, 
where (differently from   Ref.~\cite{DiLuzio:2017tjx})  a non-vanishing hypercharge $\Y^{a}$
is assigned to the $\Psi_{L,R}^{a}$ fields. 
As we will show below, this choice is crucial for addressing the problem of stable fractionally-charged relics 
in the post-inflationary PQ breaking scenario.
However, we need to ensure that  in spite of the contribution of the new fields  charged under the SM gauge group
to the running of the gauge couplings, the model  remains free from low-scale Landau poles  
below the PQ-breaking scale $\Lambda_{\rm PQ}$.\footnote{We note 
that the field content of
Ref.~\cite{Lu:2023ayc} can be embedded in complete representations of the
trinification group $\SU(3)_C \times \SU(3)_L \times \SU(3)_R$, which helps
to tame the $\U(1)_Y$ Landau pole.  The $\SU(3)_C$ Landau pole caused by the
proliferation of colored representations, however, remains. 
}

\begin{table*}[t!]
\centering 
\begin{tabular}{|c|c|c|c|c|c|c|c|}
 \hline
               &Lorentz & $\SU(3)_C$       & $\SU(2)_L$ & $\U(1)_Y$    & $G_{LR}^{(\N)} \rightarrow \SU(\N)_{L+R}$ & $\U(1)_{\rm PQ}$ & $\U(1)_{\rm \Q-\Psi}$ \\ 
               \hline
$Y$             & $(0,0)$         & $1$              & $1$       & $0$         & $(\N, \bar{\N})\rightarrow \N^2-1\oplus 1$     & $1$ & $0$ \\        
 \hline       
$\Q_L$           & $(\frac{1}{2},0)$       & $3$     & $1$       & $0$      & $(\N,1)\rightarrow \N$           & $1$ &  $1$ \\
$ \Q_R$           & $(0,\frac{1}{2})$       & $3$ & $1$       & $0$      & $(1, \N)\rightarrow \N$             & $0$ & $1$  \\
$\Psi_L^a$      & $(\frac{1}{2},0)$       & $1$              & $1$       & $\Y^{a}$      & $(\bar \N,1)\rightarrow \bar \N$   & $-1$ & $-1$   \\
$\Psi_R^a$      & $(0,\frac{1}{2})$       & $1$              & $1$       & $\Y^{a}$      & $(1,\bar \N)\rightarrow \bar \N$  & $0$ & $-1$   \\
\hline
 \end{tabular}
\caption{\label{tab:NewMatterContent} 
Field content of the $G_{LR}^{(\N)} \equiv \SU(\N)_L \times \SU(\N)_R$ model that realizes an accidental 
PQ symmetry for $\N > 4$. 
The index $a=1,2,3$ labels three copies of fermion species, with hypercharge $\Y^{a}$. 
The minimal model is obtained for $\Y^{1,2} = -1/3$ and $\Y^{3} = 2/3$.     
The last two columns display the high-quality accidental symmetries, $\U(1)_{\rm PQ}$ and $\U(1)_{\rm \Q-\Psi}$, 
broken respectively by effective operators of canonical dimension 
$d=\N$ and (at least) $d=3 \, \N /2$.
}
\end{table*}

We choose a basis in which the SM quark masses are real while
$\theta_{\rm QCD}\neq 0$.  Without loss of generality the $\Psi_{L,R}$
couplings can be taken flavor-diagonal, so that the Yukawa terms
can be written as
\begin{equation}
  \label{eq:Lm}
\mathcal{L}_{\rm Yuk} = 
- e^{i\frac{\eta_0}{\N}} y_\Q\, \bar \Q_L Y \Q_R -
   e^{i\frac{\eta_a}{\N}} y_{\Psi^a} \bar \Psi_L^a Y^\dagger \Psi_R^a \,,
\end{equation}
where $y_\Q,y_{\Psi^a}$ ($a=1,2,3$) are four real non-negative parameters.  
For the VEV 
configuration $\hat Y = \frac{1}{\sqrt N}\mathrm{diag}(1,\ldots, 1)$
all the fermions
become massive, with degenerate masses within each $\SU(3)_C$ and
$\SU(\N)_{L+R}$ multiplet.  We first show that the fermion masses
stemming from \eqn{eq:Lm} can be brought into real form without
inducing mixed $G^{(\N)}_{LR}$ anomalies.
After spontaneous $G^{(\N)}_{LR}$ breaking,
$\arg\det(M_\Q)=\eta_0+\langle a\rangle/f_a$ and
$\arg\det(M_{\Psi^a})=\eta_a-\langle a\rangle/f_a$, that is there are four independent
phases that we wish to cancel (four conditions).  We can perform four
chiral rotations of the fermion multiplets respectively with phases
$\alpha_0,\alpha_a$, subject to a fifth condition $\sum_{a=1}^3
\alpha_a =3\alpha_0$ which avoids mixed anomalies with the $\SU(\N)_{L,R}$
gauge groups. The phase of $Y$ can also be redefined (this changes the
argument of the cosine in \eqn{VD} by the addition of a constant term
$\varphi \to \tilde\varphi$).  All the complex phases can thus be
canceled. However, the chiral rotation of $\Q_{L,R}$ is anomalous with
respect to $\SU(3)_C$ of color, and  another source of explicit $\U(1)_{\rm PQ}$
breaking is then introduced.
Including the latter \cite{diCortona:2015ldu}, the relevant axion
potential takes the form
\begin{align}
  \label{eq:Vd}
  V_a &= \Delta V
  \cos\!\left(\frac{a(x)}{f_a}+\tilde\varphi\right) \nonumber \\
  &{}-m^2_\pi f^2_\pi
  \sqrt{1-\frac{4m_um_d}{(m_u+m_d)^2}
  \sin^2\!\left(\frac{a(x)}{2f_a}\right)}\, ,
\end{align}
where 
$\tilde\varphi$ is a generic constant unrelated to
$\theta_{\rm QCD}$.  We have shifted the origin of the axion field so that
$a(x)/f_a+\theta_{\rm QCD}\to a(x)/f_a$; the anomalous gluon coupling is then
$\frac{\alpha_s}{4\pi}\frac{a(x)}{f_a}G\widetilde G$.
Values of the microscopic phase of $Y$ separated by $2\pi/\mN$ are
related by a center transformation of one of the two gauge factors.
Equivalently, the faithfully acting scalar symmetry contains the
corresponding discrete quotient. Since $a/f_a=\arg\det Y$ has
physical period $2\pi$, the QCD potential has a
single gauge-inequivalent minimum and $N_{\rm DW}=1$ 
(we will come back on this point later in Sec.~\ref{sec:cosmo-defects}).

From \eq{eq:Vd} it follows that if $\kappa,\tilde\varphi=\mathcal{O}(1)$, 
then $|\langle a\rangle|/f_a\lesssim10^{-10}$ can be ensured only if the
explicit breaking satisfies 
\beq 
\label{eq:qualitybound}
\frac{\Delta V}{\chi_{\rm QCD}} \lsim 10^{-10} \, , 
\eeq
where $\chi_{\rm QCD} = \frac{m_u m_d}{(m_u + m_d)^2} m^2_\pi f^2_\pi \simeq (75.5 \, \text{MeV})^4$
and $\Delta V$ is defined in \eq{eq:DeltaV}. 
Imposing $f_a \gtrsim 10^9~\text{GeV}$ from astrophysical constraints and taking, for instance, $\Lambda_{\rm PQ} = M_{\rm Pl}$, 
the quality condition requires $\N \geq 9$.

\section{Accidental symmetries}
\label{sec:accidental-symmetries}

The renormalizable Lagrangian associated to the field content in \Table{tab:NewMatterContent} 
(for $\N > 4$)
reads 
\begin{align}
&\mathcal{L}_{(d\leq 4)} = 
-\frac{1}{4}  (\F_L)_{\mu\nu}  (\F_L)^{\mu\nu} 
-\frac{1}{4}  (\F_R)_{\mu\nu}  (\F_R)^{\mu\nu} \nonumber \\
&+ \Tr [ (D_\mu Y)^\dag (D^\mu Y) ] - V(Y^\dag Y) + \sum_{\psi = \Q, \Psi^a} \bar{\psi} i \slashed{D} \psi  \nonumber \\
&- y_\Q \bar{\Q}_L Y \Q_R - y_{\Psi^a} \bar{\Psi}^a_L Y^{\dagger}  \Psi^a_R 
+ \text{h.c.} \, ,
\end{align} 
where, without loss of generality,  the Yukawa couplings for the $\Psi^a$ fields 
have been written in the flavor diagonal  basis.
The global symmetry $K$ of the gauged kinetic terms is explicitly
broken by the generally nondegenerate Yukawa interactions
down to 
$\U(1)^5 \equiv \U(1)_{\rm PQ} \times \U(1)_{\Q} \times \U(1)_{\Psi^{1}} \times \U(1)_{\Psi^{2}} \times \U(1)_{\Psi^{3}}$, 
where a possible choice for the $\U(1)_{\rm PQ}$ charges ($Q_{\rm PQ}$) is provided in the 7th column of \Table{tab:NewMatterContent}, 
while $\U(1)_{\Q}$ and $\U(1)_{\Psi^{a}}$ denote respectively the $\Q$-number and the family-dependent $\Psi$-number. 
The $\U(1)^5$ symmetry is further broken by chiral anomalies. 
Considering a generic $\U(1)$ linear combination of the abelian factors in $\U(1)^5$, 
with generator denoted as $\alpha_{\rm PQ} Q_{\rm PQ} + \alpha_{\Q} Q_{\Q} + \sum_a \alpha_{a} Q_{\rm \Psi^a}$,  
it follows that the only non-trivial anomalies are 
$\A (\U(1) \SU(3)^2_C) \propto \alpha_{\rm PQ}$ and $\A (\U(1) \SU(\N)^2_{L,R}) \propto 3 \alpha_{\Q} + \sum_a \alpha_{a}$. 
Therefore, it is possible to identify three linearly-independent $\U(1)$'s which are non-anomalous:  
$\U(1)_{\Q-\Psi}$ (with $\alpha_{\rm PQ} = 0$ and $\alpha_{\Q} = - \alpha_1 = - \alpha_2 = - \alpha_3$), 
$\U(1)_{\Psi_1-\Psi_2}$ (with $\alpha_{\rm PQ} = \alpha_{\Q} = \alpha_3 = 0$ and $\alpha_1 = - \alpha_2 $), and 
$\U(1)_{\Psi_1-\Psi_3}$ (with $\alpha_{\rm PQ} = \alpha_{\Q} = \alpha_2 = 0$ and $\alpha_1 = - \alpha_3 = 1$), 
while $\U(1)_{\rm PQ}$ and $\U(1)_{\Q+\Psi}$ (with $\alpha_{\rm PQ} = 0$ and $\alpha_{\Q} = \alpha_1 = \alpha_2 = \alpha_3$) 
are anomalous. Note that the $\U(1)_{\Q+\Psi}$ symmetry can be employed to rotate away the linear combination 
$\theta_L + \theta_R$ (where $\theta_L$ ($\theta_R$) denotes the topological angle of $\SU(\N)_{L(R)}$), while the 
linear combination $\theta_L - \theta_R$ remains physical, possibly yielding  high-energy CP-violating 
effects.\footnote{The latter can be in principle forbidden by enforcing a $\mathbb{Z}_2$ symmetry that exchanges $L \leftrightarrow R$.} 
Crucially, the model is constructed in such a way that $\A (\U(1)_{\rm PQ} \SU(\N)^2_{L,R}) = 0$ so that the axion potential 
does not receive contributions from the confining $\SU(\N)_{L+R}$ group. 

\subsection{Quality of accidental symmetries} 

We are interested in understanding the quality of the $\U(1)^5$ symmetry, after including effective operators, whose specific form also depends on the choice of the hypercharge factors $\Y^a$ in \Table{tab:NewMatterContent}. 
The hypercharges must cancel the new gauge anomalies,
allow the dimension-six operators in Eq.~\eqref{eq:Ld6} that reduce the
separate fermion numbers to the residual $\U(1)_{\Q-\Psi}$, give integer
electric charges to the stable baryons when $\mN\equiv0\pmod{3}$, as
discussed in Sec.~\ref{sec:spectrum-baryons}, and preserve perturbativity of
$\U(1)_Y$ up to the PQ-breaking scale $\Lambda_{\rm PQ}$, as discussed in
Sec.~\ref{sec:uv-constraints}.
In the following, we focus on the minimal assignment
$\Y^{1,2}=-1/3$ and $\Y^3=2/3$.  A complete classification of viable
non-minimal assignments, together with their axion--photon couplings, is
given in \app{sec:hypercharge}.
At $d=6$, it is possible to write the following operators that break further 
the $\U(1)^5$ symmetry:  
\begin{align}
\label{eq:Ld6}
\mathcal{L}_{(d = 6)} &\supset 
\frac{a_{ij}}{\Lambda^2_{6}} ( \bar \Psi_{L,R}^i \,\Gamma \,\Psi_{L,R}^j ) 
( \bar \psi_{\rm SM}\, \Gamma \,\psi_{\rm SM} ) \nonumber \\ 
&+ \frac{b_{i}}{\Lambda^2_{6}} ( \bar \Psi_{L,R}^i\, \Gamma\, \Psi^3_{L,R} ) 
( \bar u_R \,\Gamma\, d_R ) \nonumber \\ 
&+ \frac{c_{i}}{\Lambda^2_{6}} ( \bar \Q_{L,R} \,\Gamma \, u_R ) 
( \bar \Psi_{L,R}^i \, \Gamma \, e_R ) \, ,  
\end{align}
Here $i,j=1,2$ because $\Psi^{1,2}$ have the common hypercharge
$-1/3$, and the $b_i$ operators connect this pair to the
$\Y^3=2/3$ species. The $\Gamma$'s denote collectively all the 
 $\gamma$-matrices Lorentz structures appropriate for each term, 
 and $u_R,\,d_R,\,e_R$ 
are any SM $R$-handed quark or lepton of the three generations.
The scale $\Lambda_6$ is independent of the PQ-breaking scale
$\Lambda_{\rm PQ}$ in Eq.~\eqref{VD}.  Because the operators in
Eq.~\eqref{eq:Ld6} preserve both PQ and the residual $\U(1)_{\Q-\Psi}$,
their mediators may lie below $\Lambda_{\rm PQ}$ without weakening axion
quality, provided they respect these symmetries.  We absorb the coefficients
into operator-dependent effective scales,
$\Lambda_{6,X}^{\rm eff}=\Lambda_6/\sqrt{|X|}$ for
$X\in\{a_{ij},b_i,c_i\}$.  
Below,
$\Lambda_6$ denotes a representative effective scale for the slowest relevant
transition.

This gives the breaking pattern:
\begin{align}
\U(1)^5 &\xrightarrow{a_{ij}} \U(1)_{\rm PQ} \times \U(1)_{\Q} \times \U(1)_{\Psi_1+\Psi_2} \times \U(1)_{\Psi_3}  \nonumber \\ 
&\xrightarrow{b_{i}} \U(1)_{\rm PQ} \times \U(1)_{\Q} \times \U(1)_{\Psi}  \nonumber \\ 
&\xrightarrow{c_{i}} \U(1)_{\rm PQ} \times \U(1)_{\Q-\Psi} \, . 
\end{align}
The $\SU(\N)_L \times \SU(\N)_R$ gauge symmetry implies that any operator is also invariant under its center, 
$\mathbb{Z}_{\N}^L \times \mathbb{Z}_{\N}^R$, 
whose action is defined by
\begin{equation}
\begin{aligned}
\mathbb{Z}_\N^L &: 
\begin{cases}
\Q_L \to e^{i \tfrac{2 \pi}{\N}} \Q_L\\
\Psi^a_ L \to e^{-i \tfrac{2 \pi}{\N}} \Psi^a_ L \, , \\
Y \to e^{i \tfrac{2 \pi}{\N}} Y
\end{cases}\\
\mathbb{Z}_\N^R &: 
\begin{cases}
\Q_R \to e^{i  \tfrac{2 \pi}{\N}} \Q_R\\
\Psi^a_R \to e^{-i  \tfrac{2 \pi}{\N}} \Psi^a_ R  \, . \\
Y \to e^{-i  \tfrac{2 \pi}{\N}} Y 
\end{cases}
\end{aligned}
\end{equation}
Given the choice of $\U(1)_{\rm PQ}$ in \Table{tab:NewMatterContent}, 
$\mathbb{Z}_\N^L$ is a subgroup of $\U(1)_{\rm PQ}$, 
while $\mathbb{Z}_\N^{L+R}$ is a subgroup of $\U(1)_{\Q-\Psi}$. 
Therefore, to understand the quality of the residual global symmetries $\U(1)_{\rm PQ}$ and $\U(1)_{\Q-\Psi}$, 
it is useful to look at the invariance under their discrete (gauged) subgroups $\mathbb{Z}_\N^L$ and $\mathbb{Z}_\N^{L+R}$. 
Denoting the effective number of field species by $N_\psi = n^+_\psi - n^-_\psi$, where $n^+_\psi$ ($n^-_\psi$) is the number of $\psi$ ($\psi^\dagger$) fields 
in a given gauge invariant operator, invariance under $\mathbb{Z}_\N^L$ requires
$N_{\Q_L}-N_{\Psi_L}+N_Y\equiv 0\pmod{\N}$, 
while $(N_{\Q_L} - N_{\Psi_L} + N_Y) \neq 0$ in order to break $\U(1)_{\rm PQ}$. Therefore, the lowest dimensional $\U(1)_{\rm PQ}$ 
breaking operator 
is obtained by considering only scalar fields with $N_Y = n^+_Y = \N$, that is $\det Y$. On the other hand, 
invariance under $\mathbb{Z}_\N^{L+R}$ requires
$N_\Q-N_\Psi\equiv 0\pmod{\N}$
(with $N_\Q = N_{\Q_L} + N_{\Q_R}$ and
$N_\Psi=N_{\Psi_L}+N_{\Psi_R}$),  
while $(N_\Q - N_\Psi) \neq 0$ in order to break $\U(1)_{\Q-\Psi}$. 
Therefore, the lowest dimensional $\U(1)_{\Q-\Psi}$ breaking operator 
is obtained for $N_\Q - N_\Psi = \N$, and is at least of dimension $3/2 \, \N$.  
We conclude that the quality of the $\U(1)_{\rm PQ}$ symmetry is closely tied to that of the 
$\U(1)_{\Q-\Psi}$ symmetry, implying that at large $\N$ 
(required for a robust solution to the strong CP problem) 
the lightest state charged under $\U(1)_{\Q-\Psi}$ is stable on cosmological timescales.

\section{Spectrum and bound states}
\label{sec:spectrum-bound-states}

After the breaking
$G_{LR}^{(\mN)}\times\U(1)_{\rm PQ}\to\SU(\mN)_{L+R}$ at
$v_a=\mN f_a$, the massive gauge sector contains $\mN^2-1$ vectors with
masses of $\mathcal{O}(v_a)$.  The low-energy spectrum contains the axion, the
$\mN^2-1$ massless gauge bosons of $\SU(\mN)_{L+R}$, and the fermions
$\Q$ and $\Psi^a$, with
\begin{equation}
 m_{\Q,\Psi^a}=y_{\Q,\Psi^a}\sqrt{\frac{\mN}{2}}\,f_a\,.
 \label{eq:spectrum-masses}
\end{equation}
The dimension-six operators discussed above break the separate fermion
numbers down to $\U(1)_{\Q-\Psi}$.  When kinematically allowed, they transfer
this conserved charge from heavier species to lighter ones. 
The lightest
baryon carrying this charge is therefore stable on cosmological time
scales.  Independently of the  mass hierarchy, we impose the condition 
\begin{equation}
 \mN \equiv 0\pmod 3\,.
 \label{eq:spectrum-triality}
\end{equation}
This condition is very important  since it implies that  $[\Q^\mN]$ bound states have zero $\SU(3)_C$  triality, 
and  also ensures that  the electric  charge of every
$\Psi^\mN$ baryon is an integer, thereby preventing stable
fractionally-charged relics.

\subsection{Mass orderings and decays}
\label{sec:spectrum-orderings}

We use the following orderings as representative examples.  The
constituent-level decay chains and partonic rate estimates apply directly in
the heavy-quark regime, $m_{\Q,\Psi^a}\gtrsim\Lambda_{L+R}$.  If
confinement occurs first, the same operators induce transitions among
hadrons, whose ordering depends on the spectrum of  hadron masses, including
hypercharge and hyperfine contributions.  If
\begin{equation}
 m_\Q<m_{\Psi^{1,2}}<m_{\Psi^3}\,,
 \label{eq:spectrum-Qordering}
\end{equation}
the representative decay chain is
$\Psi^3\to\Psi^i\bar d u$, followed by
$\Psi^i\to\Q u e$, and the stable constituent is $\Q$.  As an example of a
 different  ordering, we take  the $\mQ$'s as the heavier states, and the $\Psi^a$ masses  arranged 
 with an ordering analogous to the up, down, and strange quark masses
 in the SM, that is,  the  $\Y=+2/3$ state  is  lighter than
the other  two species with  $\mathcal{Y}=-1/3$
\begin{equation}
 m_{\Psi^3}<m_{\Psi^{1,2}}<m_\Q\,.
 \label{eq:spectrum-Psiordering}
\end{equation}
In this case the representative decay chain is $\Q\to\Psi^i\bar u\bar e$, followed by
$\Psi^i\to\Psi^3 d\bar u$, and the stable constituent is $\Psi^3$. 
Other orderings among the $\Psi^a$ are possible.  If a
$-1/3$-hypercharge species is the lightest, the relevant decay arrows and the
sign of the constituent mass splitting introduced below are reversed, without
changing the qualitative conclusions.

For either hierarchy, let $\Delta E$ denote the energy released in the
three-body decay.  The rates have the parametric dependence
\begin{equation}
 \Gamma_{\rm heavy}\sim
 \frac{g_{f\mN}}{8(192\pi^3)}
 \frac{(\Delta E)^5}{\Lambda_{6}^4}\,,
 \quad g_{f\mN}=\mathcal O(10^1\text{--}10^2)\,,
 \label{eq:spectrum-decay-rate}
\end{equation}
Here the relevant Wilson coefficient has been absorbed into the
operator-dependent effective scale defined below Eq.~\eqref{eq:Ld6}, while
$g_{f\mN}$ counts the accessible color, flavor, and operator channels.  After
confinement the analogous quantity $g_H$ also contains the squared hadronic
transition form factors.
The
corresponding lifetime is
\begin{equation}
 \tau_{\rm heavy}\sim 0.65
 \left(\frac{10^{11}\,\mathrm{GeV}}{\Delta E}\right)^5
 \left(\frac{\Lambda_{6}}{M_{\rm Pl}}\right)^4
 \frac{10^2}{g_{f\mN}}\,\mathrm{s}\,.
 \label{eq:spectrum-lifetime}
\end{equation}
For $\mathcal{O}(1)$ mass splittings, $\Delta E=\mathcal{O}(m_{\rm heavy})$, and
Eqs.~\eqref{eq:spectrum-decay-rate} and \eqref{eq:spectrum-lifetime} show that 
in the benchmark mass region
the heavier species can safely decay before Big Bang Nucleosynthesis (BBN). For a
small Yukawa-induced splitting, the replacement
$m_{\rm heavy}^5\to(\Delta E)^5$ displays the strong phase-space suppression.
A lower PQ-preserving transfer scale relaxes the required energy
release as $\Lambda_6^{4/5}$.  The $\mN=15$ and $18$ lifetime illustrations
below use $\Lambda_6=10^{12}\,\mathrm{GeV}$, while the $\mN=21$ benchmark in
Sec.~\ref{sec:uv-constraints} admits the common choice
$\Lambda_6=\Lambda_{\rm PQ}=10^{14}\,\mathrm{GeV}$.  In both cases the
contact scale remains above the relevant constituent and confinement scales.

A sufficiently long-lived species could instead produce an early
matter-dominated era and entropy injection~\cite{Cheek:2023fht,
DiLuzio:2024xnt}.  In all our benchmarks, however, we require every unstable
state to have an open decay channel with lifetime below $1\,\mathrm{s}$.  
A temporary matter-dominated epoch then ends before BBN, and its entropy release
can only dilute the exotic abundance, thereby relaxing the corresponding relic constraint.  
The axion DM abundance
can in principle still be modified.  For dimension-six decays, the KSVZ
analysis of Ref.~\cite{DiLuzio:2024xnt} finds that the nonstandard branch
merges with the standard axion-cosmology region when the effective suppression
scale is below approximately $0.7M_{\rm Pl}$.  Our benchmark transfer scales
lie well below this value,  
thus implying a 
standard radiation-dominated axion cosmology.

\subsection{Stable baryon candidates}
\label{sec:spectrum-baryons}

After  $\SU(\mN)_{L+R}$ confinement, mesons carry no 
$\U(1)_{\Q-\Psi}$ charge, whereas baryons made from $\mN$ fundamentals
carry the conserved charge.
The pure states are $\mB_\Q=[\Q^\mN]$ and
$\mB_{\Psi^a}=[(\Psi^a)^\mN]$, together with their charge conjugates.  More
generally, the tower with positive $\U(1)_{\Q-\Psi}$ charge can be written as
\begin{equation}
 \mathcal B_{+}(\mathbf n)=
 \biggl[\Q^{n_\Q}\prod_{a=1}^{3}(\bar\Psi^a)^{n_a}\biggr].
\end{equation}
Here $\mathbf n\equiv(n_\Q,n_1,n_2,n_3)$ is the composition vector, with
$n_\Q+\sum_{a=1}^{3}n_a=\mN$.
The opposite-charge tower is its charge conjugate.  The mass ordering
and charge-transfer decays determine which tower contains the stable endpoint.
In the ordering
\eqref{eq:spectrum-Qordering}, the stable baryon is
\begin{equation}
 \mB_\Q=[\Q^{\mN}]\,.
 \label{eq:spectrum-Qbaryon}
\end{equation}
Although $\mB_\Q$ need not be a bare QCD singlet, condition
\eqref{eq:spectrum-triality} ensures zero QCD triality.  It can then be dressed by gluons into a QCD
singlet.  Since $\Q$ has vanishing electric charge, the dressed state is
neutral. The color and spin of the lowest spatial state are fixed
by Fermi statistics.  The $\SU(\mN)_{L+R}$ Levi--Civita contraction is
antisymmetric and the $s$-wave orbital state is symmetric, so the product of
QCD color and spin must be symmetric.  Young tableaux make this condition
transparent. Under permutations of the $\mN$ identical quarks, the QCD-color and spin
wave functions furnish representations of $S_{\mN}$ acting on separate
Hilbert-space factors.  Their tensor product contains a fully symmetric
component only when the two factors carry the same Young-diagram symmetry.
The color and spin factors must therefore be associated with the same
Young-diagram shape. A bare QCD singlet would have $\mN/3$ columns of height
three, and hence a three-row permutation tableau.  It cannot be matched by
an $\SU(2)$ spin tableau, which has at most two rows.  The $s$-wave ground
state must therefore carry QCD color.  Minimizing its spin selects the
following two-row tableau, shown schematically:
\begin{equation}
 \Yboxdim{7pt}
 \begin{aligned}
 \mN\ \text{even}:\quad &
   \yng(2,2)\,\substack{\cdots\\[-1mm]\cdots}\,\yng(1,1)
   \,,\qquad J=0,
 \\[5mm]
 \mN\ \text{odd}:\quad &
   \yng(2,2)\,\substack{\cdots\\[-1mm]\cdots}\,\yng(2,1)
   \,,\qquad J=\frac12.
 \end{aligned}
\end{equation}
For even $\mN$, the first tableau gives $J=0$ and the QCD irrep
$(p,q)=(0,\mN/2)$; for odd $\mN$, the second gives $J=1/2$ and
$(p,q)=(1,(\mN-1)/2)$.  Using
$\dim(p,q)=\frac12(p+1)(q+1)(p+q+2)$, their dimensions are
\begin{equation}
 \dim R_\Q=
 \begin{cases}
  \dfrac{(\mN+2)(\mN+4)}{8}, & \mN\ \text{even},\\[2mm]
  \dfrac{(\mN+1)(\mN+5)}{4}, & \mN\ \text{odd}.
 \end{cases}
\end{equation}
Thus $\dim R_\Q=80$, $55$, and $143$ for $\mN=15$, $18$, and
$21$, respectively.
The conjugate irreps describe the corresponding antibaryons.  In both cases
the triality is $p+2q\equiv\mN\pmod{3}$; hence condition
\eqref{eq:spectrum-triality} makes it vanish and allows the colored baryon to
be screened to a singlet by gluons.

In the ordering \eqref{eq:spectrum-Psiordering}, the baryon made only from
the lightest flavor would be
\begin{equation}
 \begin{aligned}
 \mB_\Psi&=[(\Psi^3)^{\mN}]\,, &
 \Y(\mB_\Psi)=Q_{\rm em}(\mB_\Psi)&=\frac{2\mN}{3}\,,\\
 && J_{\mB_\Psi}&=\frac{\mN}{2}\,.
 \end{aligned}
 \label{eq:spectrum-Psibaryon}
\end{equation}
Here $J_{\mB_\Psi}$ is the total spin.  Its maximal value follows from the
Pauli principle: for a one-flavor spatial ground state, the antisymmetric
$\SU(\mN)_{L+R}$ singlet requires a completely symmetric spin wave function.
This is not necessarily the lightest baryon.  For
$\mN\equiv 0\pmod{3}$, the mixed-flavor tower
\begin{equation}
 \begin{aligned}
 \mB_q&=\left[(\Psi^3)^{\mN/3+q}
 (\Psi^1)^{2\mN/3-q}\right]\,,\\
 \Y(\mB_q)&=Q_{\rm em}(\mB_q)=q\,.
 \end{aligned}
 \label{eq:spectrum-Psitower}
\end{equation}
contains the neutral state
\begin{equation}
 \mB_0=\left[(\Psi^3)^{\mN/3}(\Psi^1)^{2\mN/3}\right]\,.
 \label{eq:spectrum-neutral-baryon}
\end{equation}
Since the $\Psi^a$ are $\SU(2)_L$ singlets, their electric charge
equals their hypercharge.  Writing the signed splitting
$\Delta m_{13}\equiv m_{\Psi^1}-m_{\Psi^3}$, the competition between the
constituent masses and the hypercharge self-energy gives, up to hyperfine
corrections,
\begin{equation}
 m_{\mB_q}-m_{\mB_0}\simeq
 -q\,\Delta m_{13}
 +c_Y\alpha_Y(\Lambda_{L+R})
 \Lambda_{L+R}q^2\,.
 \label{eq:spectrum-Ysplitting}
\end{equation}
Here $\alpha_Y(\mu)\equiv g_Y^2(\mu)/(4\pi)$ and $c_Y>0$ is an
$\mathcal{O}(1)$ coefficient parametrizing the baryon hypercharge distribution.
The positive sign follows from the gauge-field energy outside a composite
state: for a baryon of radius $R\sim\Lambda_{L+R}^{-1}$ and total
hypercharge $q$, this contribution is
$\delta m_Y^{\rm out}\simeq\alpha_Y q^2/(2R)>0$.
The neutral baryon is therefore selected when, parametrically,
\begin{equation}
 |\Delta m_{13}|\lesssim
 c_Y\alpha_Y(\Lambda_{L+R})\Lambda_{L+R}\,.
 \label{eq:spectrum-neutral-condition}
\end{equation}
This is analogous to the competition between quark-mass differences and
electromagnetic self-energies in QCD hadron splittings
~\cite{Gasser:2020mzy}.  Hyperfine and other short-distance effects can
modify the order-one coefficient but not the positive long-distance
contribution responsible for the mechanism.  The neutral state can also
have much lower spin than the one-flavor baryon.

Having identified the relevant hadrons, we can now state the role of the
decay epoch without ambiguity.  A heavy species that survives until
$\SU(\mN)_{L+R}$ confinement first hadronizes, after which the same
dimension-six operators induce
transitions between the corresponding hadrons.  Such flavor-changing baryon
decays occur on time scales parametrically much longer than
$\Lambda_{L+R}^{-1}$, so all thermally populated flavors participate
in hadronization even though only the lightest baryonic endpoint survives at
late times.  For generic mass splittings, these transitions leave the list of
late-time stable candidates unchanged.
In the nearly degenerate neutral branch, however, the hadronic mass ordering
is essential for the decay kinematics; we return to this point in
Sec.~\ref{sec:cosmo-Psineutral}.

Eqs.~\eqref{eq:spectrum-Qbaryon}--\eqref{eq:spectrum-neutral-condition}
identify the possible stable states.  Their cosmological viability depends
on the confinement history and on the efficiency with which baryons form,
which we discuss next.

\section{Post-inflationary cosmology}
\label{sec:postinfl-cosmology}

We now test whether the spectrum identified above admits a consistent
post-inflationary history. There are four logically distinct questions.
First, the PQ transition must produce the strings on which a single QCD
axion domain wall can end.  Second, the unstable heavy states must disappear
before BBN. Third, if a constituent is lighter than
$\Lambda_{L+R}$, the associated pNGBs must
also have efficient decay channels, rather than giving rise to an additional 
 population of stable relics.  This issue  is analyzed in
App.~\ref{app:pngb}.  Finally, the abundance of the lightest state  carrying a 
$\U(1)_{\Q-\Psi}$ charge  must respect the limits for  stable heavy relics.  The
last question depends on which type of baryonic bound states inherits the conserved charge, so
we discuss separately the two orderings given in Eqs.~\eqref{eq:spectrum-Qordering}
and \eqref{eq:spectrum-Psiordering}.

Throughout this section we assume
$\Lambda_{L+R}\gg\Lambda_{\rm QCD}$ and denote by
$T_c\simeq\Lambda_{L+R}$ the 
$\SU(\mN)_{L+R}$ confinement temperature. This  scale is generated by
dimensional transmutation, and at  one loop is given by
\begin{equation}
 \Lambda_{L+R}=v_a
 \exp\!\left[-\frac{24\pi^2}
 {(11\mN-2N_f)\,g^2_{\mathrm{L+R}}(v_a)}\right]\ll v_a\,,
 \label{eq:LambdaLR}
\end{equation}
where $N_f$ counts the Dirac fundamentals active in the running between
$v_a$ and confinement (for example $N_f=3$ if only the three QCD-color
copies of $\Q$ are active). 

We finally assume that there is no particle-antiparticle asymmetry in the heavy fermion sector, 
that is that  the Universe has an overall  vanishing  primordial 
$\U(1)_{\Q-\Psi}$ charge. This condition is preserved by the subsequent
thermal evolution: all SM fields are neutral under $\U(1)_{\Q-\Psi}$, and
the transfer operators in Eq.~\eqref{eq:Ld6} preserve it exactly.
Consequently, even if these reactions are in chemical equilibrium during
baryogenesis, they cannot convert the SM baryon or lepton asymmetry into a
net asymmetry of the stable sector.  They may redistribute chemical
potentials among $\Q$, $\Psi^a$, and the SM, but 
will always maintain a  zero
total $\Q - \Psi$ charge.\footnote{A primordial asymmetry generated by additional UV
interactions that violate $\U(1)_{\Q-\Psi}$ could  produce an
irreducible component of  stable $\mN$-baryons.  This possibility   is  
not included in our baseline cosmological framework.}

\subsection{Axion strings and domain walls}
\label{sec:cosmo-defects}

The existence of a single gauge-inequivalent minimum of the QCD-induced
axion potential, $N_{\rm DW}=1$, is necessary but is not by itself
sufficient to solve the domain-wall problem: the PQ transition must also
produce the corresponding minimal string, on which a single wall can end
~\cite{Lazarides:1982tw,Lu:2023ayc}.  Here ``minimal'' denotes a generator
of the physical vacuum manifold, rather than a bound state of several
unit-winding strings.

To identify this generator, it is important to account for the gauge-center
identifications.  The symmetry acting faithfully on the scalar order
parameter and the subgroup preserved by
$\langle Y\rangle\propto\mathbf{1}_{\mN}$ can be written as
\begin{align}
 G_V&=
 \frac{\SU(\mN)_L\times\SU(\mN)_R\times\U(1)_{\rm PQ}}
      {\mathbb Z_{\mN}\times\mathbb Z_{\mN}}\, , \\
 H_V&=\frac{\SU(\mN)_{L+R}}{\mathbb Z_{\mN}}\,.
\end{align}
Consequently, the scalar vacuum manifold
$\mathcal M_Y=G_V/H_V$ satisfies
\begin{equation}
 \pi_1(\mathcal M_Y)=\mathbb Z\,.
\end{equation}
A generator can be represented by a path for which the microscopic PQ
phase of $Y$ changes by $2\pi/\mN$, while a gauge transformation ends at
a center element of one of the two $\SU(\mN)$ factors.  The combined
configuration closes exactly, although
\begin{equation}
 \frac{\Delta a}{f_a}
 =\Delta\arg\det Y=2\pi\,.
\end{equation}
Thus the generator carries one unit of physical axion winding.  This is the
Lazarides--Shafi gauge identification anticipated in the Introduction.

The Kibble--Zurek mechanism at the $Y$ transition therefore produces a
network containing these unit-winding strings; possible higher-winding
strings are composite representatives of the same
$\pi_1(\mathcal M_Y)=\mathbb Z$ classification. Since the QCD-induced
potential is invariant under $a\to a+2\pi f_a$, one domain wall attaches to each
minimal string at the QCD epoch, and the resulting string--wall network is
unstable~\cite{Sikivie:1982qv,Lu:2023ayc}.  This topological statement does
not depend on the detailed string profile or tension.

Strictly, this discussion treats the explicit PQ-breaking contribution
from $\det Y$ as dynamically negligible during string formation, as
justified by the PQ-quality bound.  If this contribution became relevant
before the QCD epoch, it could modify the subsequent defect evolution and
axion yield, although its invariance under $a\to a+2\pi f_a$ would not generate
a stable multi-wall network.

\subsection{The confinement dynamics}
\label{sec:cosmo-confinement}

At $T\simeq T_c$, all  
$\SU(\mN)_{L+R}$ fundamentals
populating the thermal bath will  confine into 
hadrons.  Since both  $\Q$ and  $\Psi$ species can be
active,  confinement will produce mesons, multiquark clusters, and
baryons with different flavor compositions. 
The dimension-six operators can
subsequently induce decays of heavier baryons into lighter ones while
preserving the residual $\U(1)_{\Q-\Psi}$ charge. 
These decays occur on time
scales parametrically much longer than the confinement time
$1/\Lambda_{L+R}$, and hence the flavor composition of the lightest states 
is not determined at the time of  hadronization. 
Mesons carry no net $\U(1)_{\Q-\Psi}$ charge and can
eventually decay. For $m_{\Q}$ or $m_{\Psi^a}$ below
$\Lambda_{L+R}$, chiral symmetry breaking also produces
pNGBs.  An anomaly-induced decay to SM gauge bosons
is present only when the corresponding broken current has a nonvanishing
mixed anomaly with the SM gauge group.  Modes with vanishing anomaly
coefficient instead require explicit flavor breaking or another portal; when
allowed, the dimension-six operators in Eq.~\eqref{eq:Ld6} can mediate such
decays. In the light-$\Q$ regime the full color-octet multiplet
and the axial singlet decay through the QCD anomaly.  In the light-$\Psi$
regime the axial singlet and one neutral-octet direction decay through the
hypercharge anomaly, while the remaining modes can decay through 
flavor-changing channels or via mixing, whenever the corresponding path is open.  Their
classification, indicative BBN rates, and the remaining mode-by-mode checks
are given in App.~\ref{app:pngb}. The quantity relevant for the late
relic abundance is therefore the fraction of free quarks and antiquarks
ultimately converted into baryons and antibaryons, summed over all flavor
compositions. 

The  large-$\mN$ obstruction to baryon 
assembly  identified in Ref.~\cite{casimirletter} and 
 termed  \emph{Casimir bottleneck}, guarantees that the fraction of quarks that hadronize
  into baryons relative to that hadronizing into mesons is  suppressed by several orders of magnitude.
   At the beginning of the assembly chain, adding a
quark to a small cluster of $p$-quarks in the most attractive antisymmetric configuration   is controlled, 
under Casimir scaling, by an
attractive potential weighted by a relative factor of order $p$.   
Meson formation and $p$-quarks cluster destruction  
are instead controlled by an attractive potential with  relative weight factors of order  $\mN$ and
$\mN-p$, respectively~\cite{casimirletter}.  Therefore,  for
$p\ll\mN$, quarks and antiquarks  preferentially form mesons,  while the few 
small clusters that can form are efficiently destroyed  before  they can grow into an
$\mN$-constituent baryon.  In QCD ($\mN=3$) the corresponding effect is 
limited to a factor $1/2$ reduction in the  strength of the potential for di-quark  relative to meson formation,
and therefore baryon assembly  is not suppressed. However, at large $\mN$ the Casimir bottleneck 
 becomes increasingly effective, resulting in suppressions by several orders of magnitude in the ratio 
between the number of  baryons and mesons that form. 

In Ref.~\cite{casimirletter} the Casimir bottleneck mechanism is explained in a  simplified scenario 
with only one quark flavor ($n_f=1$), and restricting  the  network of cluster formation/destruction 
processes  to reactions involving a $p$-cluster interacting with a single free quark or antiquark. 
This approximation has been  termed {\it nearest-neighbor} (NN) with reference to cluster space,  
since the  only  transitions  included are $[\Q^p] \to [\Q^{p\pm 1}] $.  An extension of the  analysis  
that goes beyond the NN approximation, and also addresses the phenomenologically relevant case $n_f=3$ required in the present work,  
is presented in the more comprehensive paper~\cite{inprogress},  companion to   Ref.~\cite{casimirletter}.
Going beyond the NN approximation (BNN) by  including   cluster-cluster and cluster-anticluster reactions, 
opens  up  additional path towards large clusters formation. With respect to the NN approximation, the baryon yield 
in BNN  is enhanced by factors of $\mathcal{O}(10)$, which do not  spoil the strong large-$\mN$ Casimir suppression. 
Conversely, the extension  to $n_f=3$ brings in spin-flavor effects that contribute  additional suppression factors 
of a few orders of magnitude with respect to the  $n_f=1$ case. 

  Below, we use the  baryon densities  resulting from integrating the BNN network 
 of cluster reactions  as an indicative estimate for the baryon yield resulting from confinement.
 It is understood that the numerical results should not be taken as precise predictions for the 
  relic densities, although it is reasonable to expect  that   the Casimir  bottleneck  suppression
 will   survive in any treatment of the hadronization process in  large $\mN$ models.
 A description of the BNN  network of reactions, the spin--flavor factors,
normalization of the Casimir coefficients, the numerical pseudo-yields, and the assumptions required 
to match them onto a final relic population of baryons are discussed in detail in Ref.~\cite{inprogress}.
For convenience, the main results are collected in App.~\ref{app:confinement-hadronization}.
In solving the equations for the network of reactions, we  normalize the $p$-clusters 
and the baryon yields to $\Lambda^3_{L+R}$   rather than to the  entropy density.
For example 
\begin{equation}  
y_\mB = \frac{n_\mB}{\Lambda^3_{L+R}}\,,
\label{eq:pseudoB}
\end{equation}
represents the one-sector baryon pseudo-yield. 
We use pseudo-yields, rather than  yields normalized to  entropy density
because pseudo-yields    directly provide an approximate estimate of  the 
baryon to meson number density ratio  $n_\mB/n_\mM$, given that at temperatures 
of order $T_c\sim\Lambda_{L+R}$,   $n_{\Psi,\bar\Psi} \approx \Lambda_{L+R}^3$, and 
essentially all quarks and antiquarks   recombine into mesons.
The pseudo-yields obtained from the hadronization model  can be 
straightforwardly  converted  into a  total baryon cosmological abundance 
\begin{equation}
 \begin{aligned}
 Y_{\mB}&\equiv\frac{n_{\mB}+n_{\bar\mB}}{s}
 =\frac{45}{2\pi^2g_{*s}(T_c)}\, 2 \,y_{\mB}\,,\\
 \end{aligned}
 \label{eq:cosmo-yield}
\end{equation}
where, 
under the assumption of charge-symmetric initial
conditions $y_1(0)=\bar y_1(0)$,  $2 y_\mB=y_\mB + y_{\bar\mB} $. 
%
%
If no significant $\mB-\bar\mB$  annihilation occurs after hadronization, 
the present contribution to the cosmological energy density 
of a baryon  of mass $m_{\mB}$ is
\begin{equation}
 \Omega_{\mB}h^2\simeq
 2.74\times10^{-2}
 \left(\frac{m_{\mB}}{100\,\mathrm{TeV}}\right)
 \left(\frac{Y_{\mB}}{10^{-15}}\right)\,,
\label{eq:cosmo-Omega}
\end{equation}
where we used $\rho_{\mB,0}=m_{\mB}Y_{\mB}s_0$, with $s_0$ the present
entropy density, and
$s_0/(\rho_c/h^2)=2.74\times10^8~\mathrm{GeV}^{-1}$.
For a specified post-confinement baryon yield, we use
Eq.~\eqref{eq:cosmo-Omega} as an upper estimate: post-confinement
annihilations can only lower the symmetric abundance.  Since the
axion is intended to remain the leading DM
component, throughout this section we require any stable exotic baryon to
contribute at most ten percent of the observed abundance,
$\Omega_{\mB}h^2\leq0.1\,\Omega_{\rm DM}h^2\simeq0.012$.
This ten-percent fraction is a benchmark defining an
axion-dominated scenario, not a universal cosmological limit. We denote the
corresponding mass-dependent yield by $Y_{10\%}(m_{\mB})$.

\subsubsection{Nature of the
\texorpdfstring{$\SU(\mN)_{L+R}$}{SU(N)L+R} confinement transition}

The quantitative relic prediction depends on the nature of the
confinement transition.  For a crossover, hadronization occurs in a homogeneous plasma,  with no bubble formation or other spatial inhomogeneities, 
and the amount of baryons that will form will then suffer the strong suppression from 
the   Casimir bottleneck.
However, if  the transition is first order,  bubbles of the confined phase will form and expand, 
while  pockets of the deconfined phase will shrink,  compressing the constituents and speeding up their hadronization~\cite{Witten:1984rs}. 
Meson formation and their subsequent decay will remove the symmetric component inside the pocket. 
However, for an initial number $N$  $(\bar N)$ of quarks (antiquarks) inside a given pocket, statistical fluctuations  imply that     
an asymmetry  $|N-\bar N| \sim \sqrt{N}$ will be generically present~\cite{Asadi:2021yml,Asadi:2021pwo,Gouttenoire:2023roe}. 
Even if the symmetric component completely annihilates, this asymmetric component survives, and the corresponding 
net global charge can only be stored in baryons or antibaryons. When homogeneous
baryon formation is strongly Casimir suppressed, the  contribution to the 
baryon relic density  from locally asymmetric quark populations, which is not captured by 
the homogeneous Boltzmann network,    can dominate the final abundance.
%
%
Pure Yang--Mills theories with $\mN\geq 3$ have a strongly first-order transition
~\cite{Svetitsky:1982gs,Lucini:2005vg,Datta:2009jn,Lucini:2012gg}.  The same
behavior is expected when all fundamental fermions are much heavier than
$\Lambda_{L+R}$ and decouple from the transition dynamics.
There are nevertheless concrete reasons for considering a crossover when
several fundamental flavors are light. 
Dynamical fundamental fermions explicitly break the center symmetry probed by
the Polyakov loop and can change the nature of the phase  
transition.  Lattice studies of $n_f=1$ and $n_f=2$ QCD find that a crossover
can persist even when the fermion masses are several times the confining scale~\cite{Alexandrou:1998wv,Saito:2011fs}.  Studies of $\SU(4)$ with several
degenerate fundamentals similarly find a crossover when their masses do not 
exceed a few times the confinement scale~\cite{LatticeStrongDynamics:2020jwi,DeGrand:2021zjw,
Ayyar:2018ppa,DeGrand:2018tzn}.  
In the $m_\Q<m_{\Psi}$ scenario, the three QCD
colors act as three flavors from the viewpoint of $\SU(\mN)_{L+R}$.  In the $m_\Psi <m_{\Q}$ 
case, the small flavor mass splittings keep the three $\Psi^a$ simultaneously
 thermally active.    Although no lattice determination is available for the large values of
$\N$ considered here,  provided the mass of the lightest triplet of  states 
  is not far above $\Lambda_{L+R}$, the possibility of a homogeneous 
  crossover remains plausible in the phenomenologically relevant 
  light multi-flavor  constructions. 
  Accordingly,  in the remainder of this paper we assume that hadronization 
  proceeds via  a homogeneous crossover, with no 
  relevant effects from spatial   inhomogeneities or asymmetries 
  from  local  statistical  fluctuations.

\subsection{\texorpdfstring{$m_\Q<m_{\Psi^a}$:  a colored lightest constituent}{mQ < mPsi}} 
\label{sec:cosmo-Qlight}

For the ordering in Eq.~\eqref{eq:spectrum-Qordering}, the decay chain leaves
the stable charge in $\Q$, which has vanishing hypercharge but transforms as
a QCD triplet.  After the heavier hadrons decay, the stable endpoint lies in
the $\mB_\Q$ sector of Eq.~\eqref{eq:spectrum-Qbaryon}.

At $\SU(\mN)_{L+R}$ confinement, $\Q\bar\Q$ mesons and
$\Q^{\mN}$ baryons form in several QCD representations.  The mesons carry no
net $\U(1)_{\Q-\Psi}$ charge and are therefore unstable.  Depending on the
spectrum, they can annihilate into QCD gluons or $\SU(\mN)_{L+R}$ gaugeballs,
or cascade to lighter unstable mesons.  Gaugeballs populated in these decays
can subsequently transfer their energy to the SM through loops of $\Q$ or
$\Psi^a$, which carry both $\SU(\mN)_{L+R}$ and SM gauge charges.  In the
heavy-fermion regime these loops are described by mixed Euler--Heisenberg
operators.  We
require both the longest-lived unstable meson and the lightest hidden
gaugeball to decay before BBN,
$\tau_{\mM},\tau_G\ll1\,\mathrm{s}$, where $\tau_{\mM}$ and $\tau_G$ denote
their respective lifetimes;
in the light-$\Q$ chiral regime, for example, color-octet
pNGBs have the indicative anomaly rate
$\Gamma_{\pi_8\to gg}\sim(\alpha_s^2/\pi^3)\Lambda_{L+R}$.
App.~\ref{app:pngb} shows that every member of this octet, as well
as the axial singlet, has a nonzero QCD-anomaly decay.
For $m_\Q\gg\Lambda_{L+R}$, integrating out the $\Q$, which is charged under
both hidden color and QCD, gives the dimension-eight Euler--Heisenberg
interaction~\cite{Juknevich:2009ji}
\begin{equation}
 \begin{aligned}
 \mathcal L_{\rm EH}&\supset
 \frac{c_{\rm EH}}{m_\Q^4}
 \operatorname{Tr}F_{L+R}^2\,
 \operatorname{Tr}G_c^2,\\[-2pt]
 c_{\rm EH}&=\mathcal O(\!\alpha_{L+R}\alpha_s\!),
 \end{aligned}
\end{equation}
where other Lorentz contractions have been suppressed.  Matching the hidden
field-strength bilinear onto the lightest $0^{++}$ gaugeball gives
$\Gamma_{G\to gg}\sim c_{\rm EH}^2\Lambda_{L+R}^9/m_\Q^8$, up to group,
phase-space, and nonperturbative matrix-element factors.  The loop channel
therefore corresponds parametrically to
\begin{equation}
 \tau_{G\to gg}\sim10^{-30}\,\mathrm{s}\,c_{\rm EH}^{-2}
 \left(\frac{10^3\,\mathrm{TeV}}{m_\Q}\right)
 \left(\frac{m_\Q}{\Lambda_{L+R}}\right)^9 ,
\end{equation}
and is safely before BBN for the representative ratio
$m_\Q/\Lambda_{L+R}\sim10$ and perturbative loop coefficients.  A much
weaker gravitational channel into two gravitons has the schematic rate
$\Gamma_{G\to 2\,{\rm grav}}\sim\Lambda_{L+R}^5/M_{\rm Pl}^4$.

We next match hidden confinement onto the abundance of the stable
baryon.  From the viewpoint of $\SU(\mN)_{L+R}$, the three QCD colors of
$\Q$ are three constituent flavors.  The relevant input is therefore the
three-flavor cluster-reaction network described in
App.~\ref{app:confinement-hadronization}.  The benchmark including all
reactions modeled there is labeled ``BNN'' in
Table~\ref{tab:cosmo-network-yields}.  The Casimir bottleneck identified in
Ref.~\cite{casimirletter} suppresses baryon assembly, while the additional
reactions included in the network of Ref.~\cite{inprogress} partially
enhance it.  The subsequent flavor cascade feeds the lightest $\mB_\Q$
state while preserving the residual conserved charge.

Under the assumed crossover and network-to-relic matching, we
denote the resulting post-confinement baryon-plus-antibaryon yield by
$Y_{\mB_\Q}$.  Using Eq.~\eqref{eq:cosmo-yield}, the corresponding
three-flavor entries of Table~\ref{tab:cosmo-network-yields} give
\begin{equation}
 Y_{\mB_\Q}=
 \left(\frac{100}{g_{*s}(T_c)}\right)
 \begin{cases}
  1.5\times10^{-20}, & \mN=15,\\
  2.6\times10^{-26}, & \mN=18,\\
  2.0\times10^{-32}, & \mN=21.
 \end{cases}
 \label{eq:cosmo-Qnetwork-yields}
\end{equation}
The assumptions entering this conversion, including the residual
hadronization uncertainty, are discussed in
App.~\ref{app:confinement-hadronization}.

Requiring the stable endpoint to contribute at most ten percent of
DM gives
\begin{equation}
 Y_{\mB_\Q}
 \lesssim Y_{10\%}(m_{\mB_\Q})
 \simeq 4.4\times10^{-16}
 \left(\frac{100\,\mathrm{TeV}}{m_{\mB_\Q}}\right),
 \label{eq:cosmo-Qbound}
\end{equation}
or, equivalently,
\begin{equation}
 \begin{aligned}
 m_{\mB_\Q}&\lesssim
 \left(\frac{g_{*s}(T_c)}{100}\right) 
 \begin{cases}
  2.9\times10^9\,\mathrm{GeV}, & \mN=15,\\
  1.7\times10^{15}\,\mathrm{GeV}, & \mN=18,\\
  2.1\times10^{21}\,\mathrm{GeV}, & \mN=21.
 \end{cases}
 \end{aligned}
 \label{eq:cosmo-Qmass-ceilings}
\end{equation}
Thus, within the stated network-to-relic matching, $\mN=15$ permits
a stable baryon up to about $3\times10^9\,\mathrm{GeV}$, while the stronger
Casimir bottlenecks at $\mN=18$ and $21$ make the  limits on the mass 
of the relic baryons  progressively
weaker.  Possible later QCD annihilation is not included in
Eqs.~\eqref{eq:cosmo-Qnetwork-yields} and
\eqref{eq:cosmo-Qmass-ceilings}.
However, especially  at the largest values of the baryon mass, 
 the number densities are so suppressed that  an additional reduction 
from annihilation is most  likely completely negligible.

\subsection{\texorpdfstring{$m_{\Psi^3}<m_\Q$}{mPsi3 < mQ}:
a hypercharged color-singlet constituent}
\label{sec:cosmo-Psilight}

For the ordering in Eq.~\eqref{eq:spectrum-Psiordering}, the stable
charge is carried before confinement by the QCD-singlet $\Psi^3$, whose
hypercharge is $\Y^3=2/3$.  At $\SU(\mN)_{L+R}$ confinement,
$\Psi\bar\Psi$ mesons and $\Psi^\mN$ baryons form.  The mesons carry no
$\U(1)_{\Q-\Psi}$ charge and are therefore unstable: they can annihilate
into pairs of hypercharge gauge bosons or cascade to lighter hidden mesons.
The latter can decay through the hypercharge portal when allowed, or through
the dimension-six flavor operators suppressed by $\Lambda_6$.  Hidden
gaugeballs populated in these processes can transfer their energy to the SM
through $\Psi^a$ loops, described in the heavy-fermion regime by the
hypercharge analogue of the mixed Euler--Heisenberg operator discussed
above.  As in the $\Q$-light branch, the longest-lived
unstable meson and the lightest gaugeball must decay before BBN.
The potentially slow pNGB modes and the axial singlet
are treated explicitly in App.~\ref{app:pngb}.

The stable baryonic endpoint can be either the charged
single-flavor state $\mB_\Psi$ or the neutral mixed-flavor state $\mB_0$
identified in Sec.~\ref{sec:spectrum-baryons}. 
The latter avoids the charged-relic bound and is therefore the most favorable candidate, on which we focus below.

\subsubsection{Hypercharge selection of a neutral baryon}
\label{sec:cosmo-Psineutral}

When Eq.~\eqref{eq:spectrum-neutral-condition} holds, the hypercharge
contribution selects $\mB_0$ as the lightest baryon in the tower of
Eq.~\eqref{eq:spectrum-Psitower}.  This is a condition on the Yukawa
splitting, rather than an additional cosmological assumption.

The same near degeneracy also keeps $\Psi^1$ thermally populated at
confinement.  Taking $T_c\simeq\Lambda_{L+R}$, the neutral-selection
condition implies $|\Delta m_{13}|=\mathcal{O}(\alpha_Y(T_c)T_c)$ or smaller, up to the
order-one coefficient $c_Y$, so the equilibrium abundance of $\Psi^1$ is not
Boltzmann suppressed relative to that of $\Psi^3$.
Moreover, the partonic decay $\Psi^1\to\Psi^3 d\bar u$ is phase-space
suppressed, with
\begin{equation}
 \Gamma_{\Psi^1\to\Psi^3 d\bar u}
 \propto
 \frac{|\Delta m_{13}|^5}
 {(\Lambda_{6,b_1}^{\rm eff})^4}\,.
 \label{eq:cosmo-Psi1-degeneracy}
\end{equation}
A population of $\Psi^1$ can therefore survive until confinement and
participate in the formation of mixed-flavor baryons.  Heavier baryons then
decay toward the lightest state.  Inside $\mB_0$, converting one $\Psi^1$
into $\Psi^3$ would produce $\mB_{+1}$.  The transition
$\mB_0\to\mB_{+1}+d\bar u$ is kinematically forbidden precisely when the
hypercharge contribution makes $\mB_0$ the lightest baryon.  Thus the $\Psi^1$
constituents are stable inside $\mB_0$, even though
$m_{\Psi^1}>m_{\Psi^3}$.
The same small splitting would lead to a lifetime problem if the
flavor-transfer operators were suppressed by the much higher PQ-breaking
scale.  With the independent scale $\Lambda_6$, Eq.~\eqref{eq:spectrum-lifetime}
instead gives the sufficient condition
\begin{equation}
 \Delta E\gtrsim9.2\times10^{10}\,\mathrm{GeV}
 \left(\frac{\Lambda_6}{M_{\rm Pl}}\right)^{4/5}
 \left(\frac{10^2}{g_H}\right)^{1/5},
 \label{eq:cosmo-neutral-BBN}
\end{equation}
where $\Delta E$ is the partonic or hadronic energy release as appropriate,
and $g_H$ denotes the corresponding channel multiplicity, including hadronic
form factors after confinement as described below
Eq.~\eqref{eq:spectrum-decay-rate}.
For the illustrative choice $\Lambda_6=10^{12}\,\mathrm{GeV}$ this becomes
only
\begin{equation}
 \Delta E\gtrsim2.0\times10^5\,\mathrm{GeV}
 \left(\frac{10^2}{g_H}\right)^{1/5}.
 \label{eq:cosmo-neutral-BBN-benchmark}
\end{equation}
If hidden confinement occurs first, the same operator induces hadronic
transitions $H_i(\Psi^1)\to H_f(\Psi^3)+d\bar u$.  Their phase space is
controlled by $\Delta E_H=M_{H_i}-M_{H_f}$, including hypercharge, hyperfine,
and binding contributions.  
The condition applies only to the unstable hadrons, and not to 
 $\mB_0$ that is cosmologically stable. Within the
baryon tower, an adjacent charged excitation has the indicative gap
\begin{equation}
 \Delta E_H^{\rm min}\simeq
 c_Y\alpha_Y(\Lambda_{L+R})\Lambda_{L+R}-|\Delta m_{13}|\,.
\end{equation}
Define the fractional position within the neutral window by
$\epsilon\equiv|\Delta m_{13}|/
[c_Y\alpha_Y(\Lambda_{L+R})\Lambda_{L+R}]<1$.  Then
Eqs.~\eqref{eq:spectrum-neutral-condition} and
\eqref{eq:cosmo-neutral-BBN-benchmark} are simultaneously satisfied for
\begin{equation}
 \Lambda_{L+R}\gtrsim
 \frac{2.0\times10^7\,\mathrm{GeV}}{c_Y(1-\epsilon)}
 \left(\frac{10^{-2}}{\alpha_Y}\right)
 \left(\frac{10^2}{g_H}\right)^{1/5}.
 \label{eq:cosmo-neutral-confinement-floor}
\end{equation}
For $c_Y(1-\epsilon)=\mathcal{O}(1)$ this lower scale is compatible with the relic
ceilings in Eq.~\eqref{eq:cosmo-Qmass-ceilings}.  For example,
$\Lambda_{L+R}\sim10^8\,\mathrm{GeV}$ at $\mN=15$ and
$10^9\,\mathrm{GeV}$ at $\mN=18$ give the confinement-dominated estimates
$m_{\mB}\sim1.5\times10^9\,\mathrm{GeV}$ and
$1.8\times10^{10}\,\mathrm{GeV}$, respectively, both below the corresponding
ten-percent DM ceilings for order-one baryon-mass coefficients.  They also
satisfy $m_{\rm light}<\Lambda_{L+R}\ll\Lambda_6$.  
Separating $\Lambda_6$
from $\Lambda_{\rm PQ}$ therefore removes the phase-space-suppressed BBN rate
obstruction without altering PQ-quality.
For the common-scale $\mN=21$ point introduced in
Sec.~\ref{sec:uv-constraints},
$f_a=10^{10}\,\mathrm{GeV}$ and $y_{\rm light}=0.1$ give
$m_{\rm light}=3.2\times10^9\,\mathrm{GeV}$, while
$v_a=\mN f_a=2.1\times10^{11}\,\mathrm{GeV}$.  We take the 
confinement benchmark
$\Lambda_{L+R}=2\times10^{10}\,\mathrm{GeV}$, for which
$\Lambda_{L+R}/v_a\simeq0.095$.  Eq.~\eqref{eq:LambdaLR} then corresponds
to $g_{L+R}(v_a)\simeq0.7$, with only a mild dependence on the fermion
thresholds, and the hierarchy
$m_{\rm light}<\Lambda_{L+R}<m_{\rm heavy}$ is manifest.  The resulting
confinement-dominated estimate
$m_{\mB}\sim4.2\times10^{11}\,\mathrm{GeV}$ remains far below the corresponding
relic ceiling.  With $\Lambda_6=10^{14}\,\mathrm{GeV}$,
Eq.~\eqref{eq:cosmo-neutral-BBN} requires only
$\Delta E\gtrsim8.0\times10^6\,\mathrm{GeV}$ for $g_H=100$, comfortably
below the indicative hypercharge splitting
$\alpha_Y\Lambda_{L+R}\sim2\times10^8\,\mathrm{GeV}$ for
$\alpha_Y\sim10^{-2}$.

This neutral state is also a singlet of QCD.  Its radius is set by
$\Lambda_{L+R}^{-1}$ rather than by
$\Lambda_{\rm QCD}^{-1}$; it therefore neither acquires a QCD-sized cloud at ordinary-QCD confinement,
as $\mB_\Q$ does, nor requires the late electromagnetic annihilation relevant
to the charged $\mB_\Psi$ branch.  $\mB_0$  is therefore the cleanest relic candidate. 
Moreover, the availability of more than one constituent
flavor allows a lower-spin ground state than for the single-flavor baryon.
Consequently $\mB_0$ interactions with SM matter arise only through
higher multipoles and polarizabilities and are strongly suppressed by the
large compositeness scale.

At confinement, the $\Psi^a$ provide the three
constituent flavors of the network described in App.~\ref{app:confinement-hadronization},   
under the assumption that the mass splittings are  sufficiently small for the
$\SU(6)$ spin--flavor weights to apply, 
and that all three $\Psi^a$ degrees of freedom are thermally populated 
with   comparable abundances.

Under the assumed crossover and network-to-relic matching,
Eq.~\eqref{eq:cosmo-Qnetwork-yields} applies with
$Y_{\mB_\Q}\to Y_{\mB_0}$.  The ten-percent DM bound and the corresponding
mass ceilings in Eqs.~\eqref{eq:cosmo-Qbound} and
\eqref{eq:cosmo-Qmass-ceilings} likewise apply with
$m_{\mB_\Q}\to m_{\mB_0}$. For the displayed normalization,
$\mN=15$ permits the illustrative neutral-baryon mass above, while the much
smaller $\mN=18$ and $21$ yields leave increasingly broad mass margins.  

\subsection{String breaking}
\label{sec:string-breaking}

The leading  effect that is not included in the hadronization network, for both 
$\chi = \Psi^a$ and $ \mQ$,  
is dynamical string breaking, which is responsible for confining into color singlets 
the residual colored clusters that survive in  non-negligible abundance after  
the Casimir network of interactions has quenched. 
In fact, when the number density of clusters becomes  sufficiently 
depleted, the network interactions freeze out, in a way analogous to the usual freeze-out
of DM annihilation.  This corresponds to  the long range confining regime in which 
hadronization  is controlled by color string breaking.  In this regime the 
$\mN$-ality of  the cluster-(anti)cluster configuration is the relevant quantity.
 Configurations with zero total $\mN$-ality can be fully screened into singlet
hadrons by quark-antiquark pairs  from the vacuum. 
For baryon production, there is only one  relevant configuration:   the only one  with  
zero $\mN$-ality that can form a baryon  by   creating   a single quark-antiquark pair from the vacuum, 
  $[\chi^{\mN-1}] + \chi \to \mB + \mM$. 
 Since the rate of this process is proportional to $y_{\mN-1}\cdot y_1$, we expect it to provide only a negligible 
contribution to the relic baryon density, while string breaking will predominantly lead to meson formation~\cite{casimirletter,inprogress}.




\section{Landau poles vs.~quality constraints} 
\label{sec:uv-constraints}

The mass orderings and relic requirements identified above must ultimately be
compatible with the PQ-quality and perturbativity conditions that motivated the construction.  We now
turn to the combined PQ-quality and Landau-pole constraints and determine
whether they overlap the cosmologically viable regions.

The choice of $\mN=15$, $18$, and $21$ follows from this combined
test.  Integer electric charge requires $\mN\equiv 0\pmod{3}$.  At
$f_a=2\times10^{11}\,\GeV$ and $2|k_D|=1$, the $\mN=12$ quality bound already
requires $\Lambda_{\rm PQ}\gtrsim2.2\times10^{19}\,\GeV>M_{\rm Pl}$.
For 
$g_{*s}(T_c)=100$, its three-flavor network yield
also restricts the stable-baryon mass to
$m_{\mB}\lesssim1.2\times10^4\,\GeV$ at the ten-percent-DM level; see
Table~\ref{tab:cosmo-network-yields}.  Smaller values of $\mN$ weaken both the
determinant suppression and the Casimir bottleneck.  Thus $\mN\leq12$ does not
provide a simultaneous solution under the baseline assumptions, although a
lower $f_a$ or additional depletion can reopen part of this parameter space.
We display $\mN=15$, the first clean benchmark, $\mN=18$, and $\mN=21$, for
which the relic suppression is strong enough to raise the lighter Yukawa to
$\mathcal{O}(0.1)$ and to identify $\Lambda_6$ with $\Lambda_{\rm PQ}$.  Values above
$\mN=21$ would further suppress PQ violation and baryon production, but do
not open a qualitatively new phenomenological regime and worsen the SM
running; we therefore do not pursue them.

\begin{figure*}[t!]
  \centering
  \begin{minipage}[t]{0.315\textwidth}
    \centering
    \includegraphics[width=\linewidth]{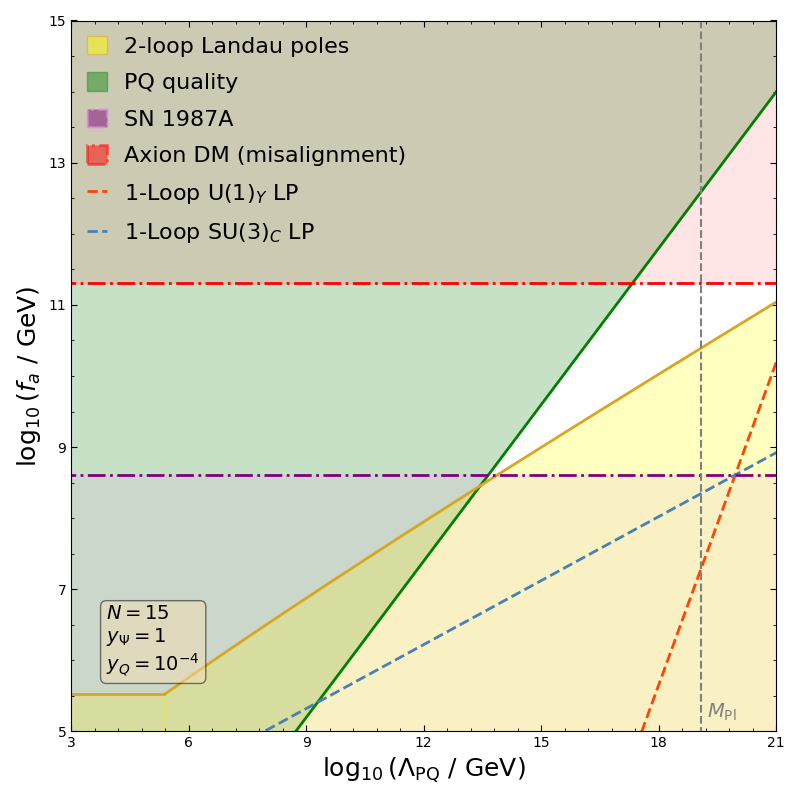}\\[-1mm]
    \textbf{(a)} $\N=15$, $m_\Q<m_\Psi$
  \end{minipage}\hfill
  \begin{minipage}[t]{0.315\textwidth}
    \centering
    \includegraphics[width=\linewidth]{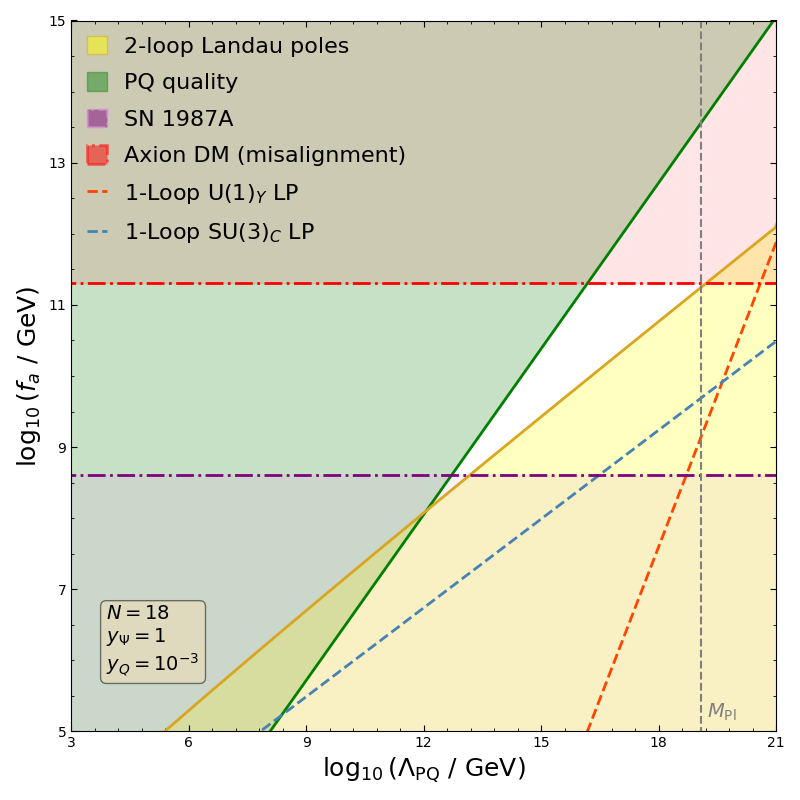}\\[-1mm]
    \textbf{(b)} $\N=18$, $m_\Q<m_\Psi$
  \end{minipage}\hfill
  \begin{minipage}[t]{0.315\textwidth}
    \centering
    \includegraphics[width=\linewidth]{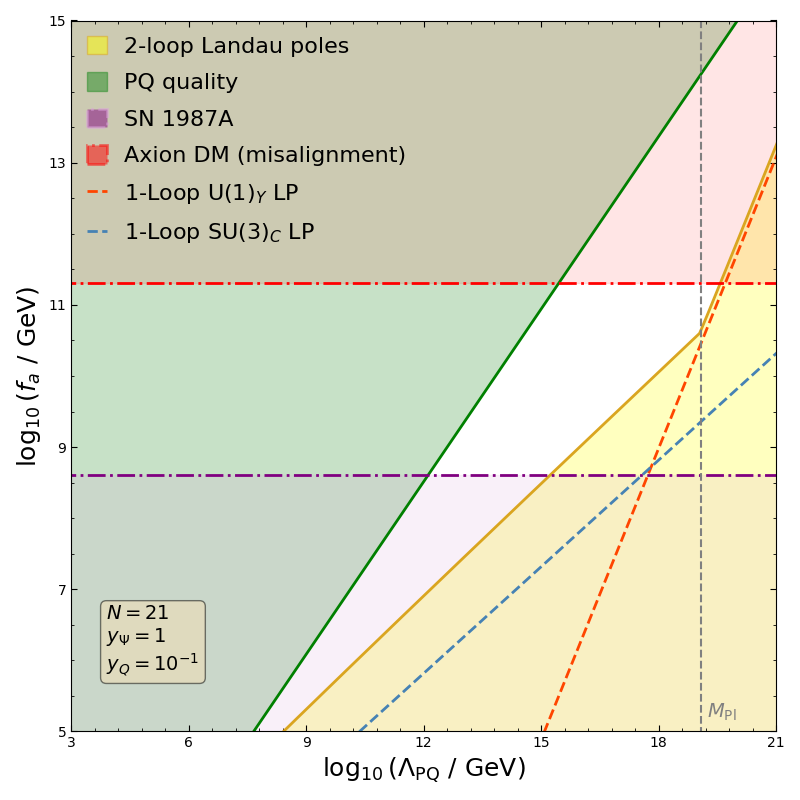}\\[-1mm]
    \textbf{(c)} $\N=21$, $m_\Q<m_\Psi$
  \end{minipage}

  \vspace{2mm}
  \begin{minipage}[t]{0.315\textwidth}
    \centering
    \includegraphics[width=\linewidth]{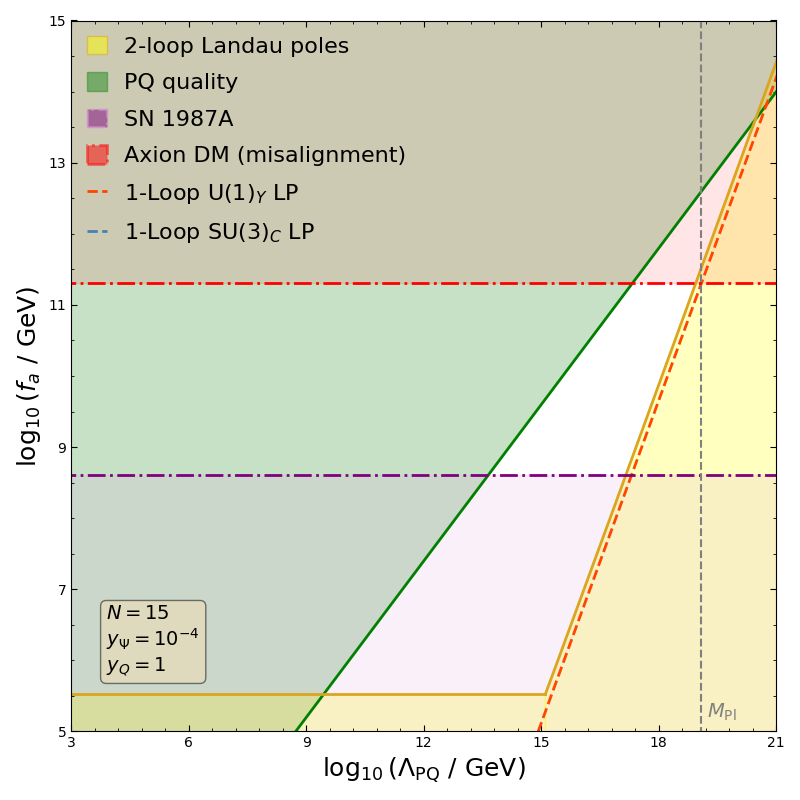}\\[-1mm]
    \textbf{(d)} $\N=15$, $m_\Psi<m_\Q$
  \end{minipage}\hfill
  \begin{minipage}[t]{0.315\textwidth}
    \centering
    \includegraphics[width=\linewidth]{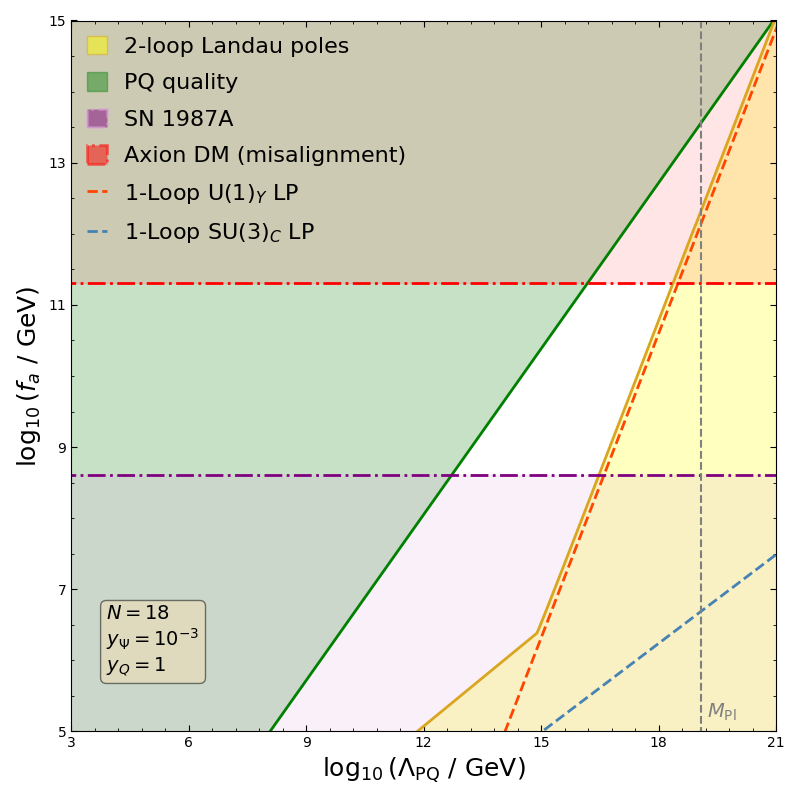}\\[-1mm]
    \textbf{(e)} $\N=18$, $m_\Psi<m_\Q$
  \end{minipage}\hfill
  \begin{minipage}[t]{0.315\textwidth}
    \centering
    \includegraphics[width=\linewidth]{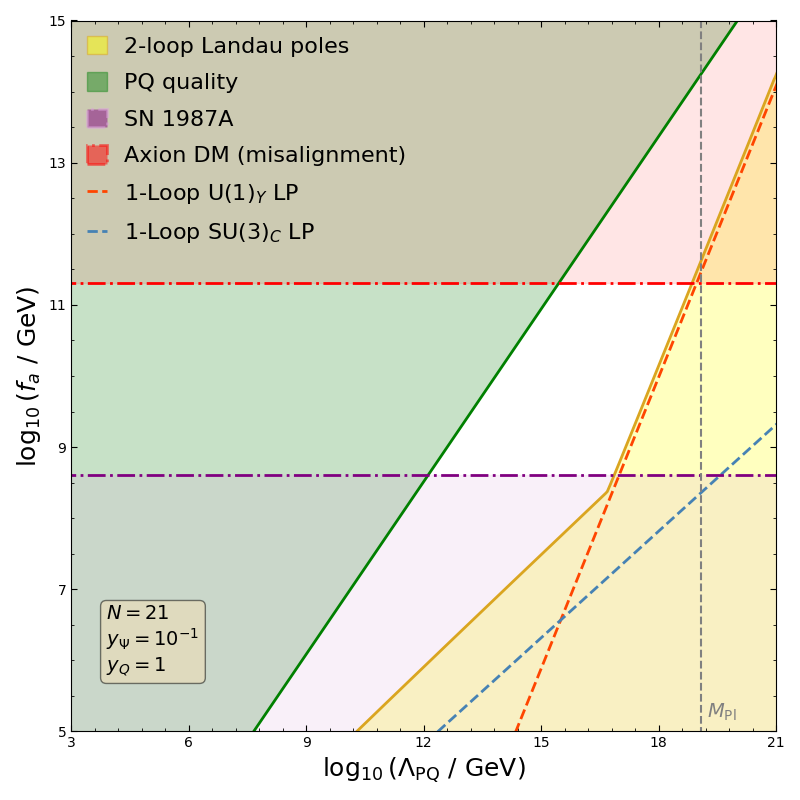}\\[-1mm]
    \textbf{(f)} $\N=21$, $m_\Psi<m_\Q$
  \end{minipage}
  \caption{PQ-quality, axion, and SM-perturbativity constraints in
  the $(\Lambda_{\rm PQ},f_a)$ plane for $\N=15$, $18$, and $21$ from left
  to right.  The top and bottom rows show $m_\Q<m_\Psi$ and
  $m_\Psi<m_\Q$, respectively.  We take $y_{\rm heavy}=1$, with
  $y_{\rm light}=10^{-4}$, $10^{-3}$, and $10^{-1}$ for $\N=15$, $18$, and
  $21$.  Green and yellow regions are excluded respectively by the
  determinant PQ-quality bound and by a two-loop SM Landau pole below
  $\Lambda_{\rm PQ}$; dashed orange and blue curves show the corresponding
  one-loop $\U(1)_Y$ and $\SU(3)_C$ boundaries.
  In panels (a) and (d), the short horizontal segment of
  the solid two-loop boundary at small $f_a$ marks $m_{\rm light}=m_Z$; below
  it the chosen Yukawa would give a new fermion lighter than the $Z$, outside
  the threshold prescription used for the plot.
  Purple shading below the
  dash-dotted SN~1987A line is excluded by stellar cooling, while red shading
  above the dash-dotted misalignment-only axion-DM line denotes axion
  overproduction.  The latter line is an upper
  reference for $f_a$ in the post-inflationary history.  The gray vertical
  line denotes $M_{\rm Pl}$.  Unshaded white regions to its left satisfy all
  displayed requirements.}
  \label{fig:paramspace}
\end{figure*}

The new fermion fields, $\Q$ and $\Psi^a$ are charged respectively under 
$\SU(3)_C$ and $\U(1)_Y$, and hence they 
could lead to Landau poles in the $g_s$ and $g_Y$ gauge couplings. 
Defining the Landau pole scale, $\Lambda_{\rm LP}$, such that $\alpha^{-1}_{s,Y} (\Lambda^{s,Y}_{\rm LP}) = 0$,   
with $\alpha_{s,Y} = g^2_{s,Y}/(4\pi)$, 
from the expression of the one-loop beta function one obtains: 
\begin{equation}\label{solNew_LP_exclusion}
\begin{aligned}
    \frac{2\pi}{\alpha_s(m_Z)}-7\ln\frac{m_Z}{\Lambda^s_{\rm LP}}+\frac{2 \N}{3}\ln\frac{m_Q}{\Lambda^s_{\rm LP}}=0 \, ,\\
    \frac{2\pi}{\alpha_Y(m_Z)}+\frac{41}{6}\ln\frac{m_Z}{\Lambda^Y_{\rm LP}}+\frac{4\N}{3}\sum_a (\Y^a)^2\ln\frac{m_\Psi}{\Lambda^Y_{\rm LP}}=0 \, ,
\end{aligned}
\end{equation}
where we have assumed a common mass threshold for the $\Psi^a$ fields and 
$\sum_a (\Y^a)^2=2/3$ in the minimal setup 
with $\Y^{1,2} = -1/3$ and $\Y^{3} = 2/3$.\footnote{Non-minimal hypercharge assignments, 
discussed in \app{sec:hypercharge}, systematically yield low-scale Landau poles in the $\U(1)_Y$ factor.}
Eq.~\eqref{solNew_LP_exclusion} is useful for displaying the threshold
dependence, but our Landau-pole analysis, including the exclusions in
Fig.~\ref{fig:paramspace}, is performed with the two-loop RG equations.  This
is particularly important for the strong coupling: its one-loop coefficient,
$a^{\rm SM+\Q}=-7+2\N/3$, is accidentally small for $\N\sim 10$, so
two-loop terms can substantially lower the scale at which perturbativity is
lost.  The quantitative example and the corresponding beta functions are
given in App.~\ref{sec:two-loop-running}.  There is no analogous cancellation
in the one-loop hypercharge coefficient, and two-loop effects are consequently
less important for the $\U(1)_Y$ Landau pole.  This difference is also visible
in the pattern of the exclusions in Fig.~\ref{fig:paramspace}.
In order for the model to be perturbatively under control until $\Lambda_{\rm PQ}$, 
the explicit PQ-breaking scale defined in \eq{VD}, we require $\Lambda^{s,Y}_{\rm LP} \geq \Lambda_{\rm PQ}$, 
otherwise one cannot claim the PQ quality problem to be solved.

These bounds have to be compared with the PQ quality bound in \eq{eq:qualitybound}, 
which reads explicitly 
\beq 
\( \frac{\N}{2} \)^{\N/2}
\( \frac{f_a^4}{\chi_{\rm QCD}} \)
\left(\frac{f_a}{\Lambda_{\rm PQ}}\right)^{\N-4} \lesssim 10^{-10} \, ,   
\eeq
where we took $2 |k_D| = 1$ in \eq{eq:DeltaV} and
used $v_a=\N f_a$ from \eq{eq:axfielddef}.

The six panels of Fig.~\ref{fig:paramspace} display the two mass
orderings for $\N=15$, $18$, and $21$.  We fix $y_{\rm heavy}=1$ and take
$y_{\rm light}=10^{-4}$, $10^{-3}$, and $10^{-1}$, respectively.  In the
light-$\Psi$ panels, $y_\Psi$ denotes the common value of the three nearly
degenerate Yukawas; the much smaller splittings that select $\mB_0$ are
irrelevant for the RG thresholds.  At $f_a=2\times10^{11}\,\GeV$, these
choices give
\begin{equation}
 m_{\rm light}\simeq
 \begin{cases}
  5.5\times10^7\,\mathrm{GeV},&\N=15,\\
  6.0\times10^8\,\mathrm{GeV},&\N=18,\\
  6.5\times10^{10}\,\mathrm{GeV},&\N=21,
 \end{cases}
 \label{eq:displayed-light-masses}
\end{equation}
while the corresponding heavy masses are $5.5$, $6.0$, and
$6.5\times10^{11}\,\mathrm{GeV}$.  The progressively stronger Casimir
bottleneck permits a sufficiently depleted baryon relic even as the lighter
Yukawa is raised to $\mathcal{O}(0.1)$ at $\N=21$.

The two rows illustrate the complementary effects of the mass
ordering.  A lighter $\Q$ lowers the colored threshold, whereas a lighter
$\Psi$ lowers the hypercharged threshold.  At the reference value of $f_a$,
the first two-loop poles occur at
$\log_{10}(\Lambda_{\rm LP}/\GeV)=21.5$, $19.2$, and $19.6$ in the
$\Q$-light $\N=15$, $18$, and $21$ panels, and at $19.0$, $18.4$, and
$18.9$ in the corresponding $\Psi$-light panels.  
The $\U(1)_Y$ pole is first in all cases except the $\Q$-light $\N=18$
benchmark, where $\SU(3)_C$ becomes limiting.  Increasing $\N$ improves PQ
protection and suppresses baryon formation, while the larger matter
multiplicity works in the opposite direction for perturbativity.  Raising
the fermion thresholds with the larger Yukawas visible in Fig.~\ref{fig:paramspace}
preserves an ample overlap.

For fixed $k_D$, the PQ-quality condition depends on $\N$, $f_a$,
and $\Lambda_{\rm PQ}$, but not on the Yukawa ordering.  The green boundary is
therefore common to the two rows at fixed $\N$.  At
$f_a=2\times10^{11}\,\GeV$, the quality floors are
$\log_{10}(\Lambda_{\rm PQ}/\GeV)=17.32$, $16.18$, and $15.44$ for
$\N=15$, $18$, and $21$.  Comparison with the two-loop poles and the gray
$M_{\rm Pl}$ line shows that both mass orderings admit simultaneous
PQ-quality and perturbativity windows for all three values of $\N$.  For $g_{*s}(T_c)=100$, the constituent-mass-dominated
benchmarks give
$\Omega_{\mB}h^2\simeq3.4\times10^{-3}$,
$7.6\times10^{-8}$, and $7.6\times10^{-12}$ for $\N=15$, $18$, and $21$,
respectively, all below the adopted ten-percent-DM value $0.012$.

The $\N=21$ benchmark also removes the need to postulate two widely
separated UV scales.  At the representative point
\begin{equation}
 f_a=10^{10}\,\GeV,\qquad
 \Lambda_{\rm PQ}=\Lambda_6=10^{14}\,\GeV,
 \label{eq:common-scale-benchmark}
\end{equation}
with $y_{\rm light}=0.1$ and $y_{\rm heavy}=1$, the quality condition only
requires $\Lambda_{\rm PQ}\gtrsim6.9\times10^{13}\,\GeV$, while the first
two-loop pole lies near $7.9\times10^{17}\,\GeV$ in either ordering.  The
constituent masses are $3.2\times10^9$ and
$3.2\times10^{10}\,\GeV$.  Together with the confinement choice discussed
in Sec.~\ref{sec:cosmo-Psineutral}, this gives 
prompt unstable-particle decays, and a stable-baryon abundance far below the
adopted limit.

The purple shading below the SN~1987A boundary denotes the stellar
cooling exclusion, represented here by
$f_a\gtrsim4\times10^8\,\GeV$~\cite{Carenza:2019pxu}.  The red shading above
$f_a=2\times10^{11}\,\GeV$ marks axion overproduction from misalignment alone
~\cite{Borsanyi:2016ksw}.  This value is an upper benchmark for axion DM in
the post-inflationary scenario: axions radiated by strings and the subsequent
string--domain-wall network increase the relic abundance, so the value of
$f_a$ giving all of DM can be lower
~\cite{Hiramatsu:2012gg,Klaer:2017ond,Gorghetto:2018myk,
Gorghetto:2020qws,Buschmann:2021sdq,Saikawa:2024bta,Benabou:2024msj}.
Accordingly, the unshaded interval between the two horizontal boundaries is
the relevant axion window shown in the figure.  All three displayed values of
$\N$ retain simultaneous PQ-quality and perturbativity regions both on the
misalignment reference and at lower $f_a$; the common-scale point in
Eq.~\eqref{eq:common-scale-benchmark} lies in the latter region.

\section{Conclusions}
\label{sec:concl}
We have presented a realization of a high-quality QCD axion
designed for a post-inflationary PQ breaking scenario.  The
$\SU(\mN)_L\times\SU(\mN)_R$ gauge structure makes $\U(1)_{\rm PQ}$
accidental and pushes its leading explicit violation to high operator
dimension.  At the same time, the minimal winding of the bifundamental field
is the axion string on which the QCD domain wall terminates.  The resulting
domain-wall number is one and the string--wall network is unstable.

The new fermions leave a high-quality $\U(1)_{\Q-\Psi}$ symmetry and hence one
stable baryonic state after $\SU(\mN)_{L+R}$ confinement.  We treated
separately the orderings $m_\Q<m_{\Psi^a}$ and 
$m_{\Psi^a}<m_\Q$.  The first
leads to an electrically neutral but QCD-dressed baryon.  In the second, if e.g.~$m_{\Psi^3} \ll m_{\Psi^{1,2}}$, the
generic lightest baryon is charged; however, when the constituent splitting is
smaller than the hypercharge self-energy, an electrically neutral
mixed-flavor QCD-singlet baryon is selected.  This neutral branch 
is the cleanest candidate.  
The PQ-preserving dimension-six scale $\Lambda_6$ can lie below the
explicit PQ-breaking scale $\Lambda_{\rm PQ}$, as illustrated by the
$\mN=15$ and $18$ lifetime benchmarks with
$\Lambda_6=10^{12}\,\mathrm{GeV}$.  More economically, the $\mN=21$ point
allows the common choice
$\Lambda_6=\Lambda_{\rm PQ}=10^{14}\,\mathrm{GeV}$ with
$y_{\rm light}=0.1$ and $y_{\rm heavy}=1$.  In either realization the
flavor-changing decays are faster than BBN.  
The pNGB and $\eta'$ analysis in App.~\ref{app:pngb} is also compatible
with BBN and with neglecting secondary baryon production.

Our relic analysis uses the Casimir bottleneck identified  in
Ref.~\cite{casimirletter} as a phenomenological model for obtaining numerical
estimates of baryon production during confinement. 
We do not solve the nonperturbative, real-time dynamics of the strongly coupled sector from first principles. 
The recombination network provides a tractable description
of this dynamics under specified assumptions, and the resulting relic
densities should be understood as model-dependent benchmark estimates.
Within this framework, the BNN network studied in detail in Ref.~\cite{inprogress} shows  that non-nearest-neighbor reactions enhance baryon formation at
large $\mN$ just by  $\mathcal{O}(10)$ factors, and do not remove its strong suppression.
On the other hand, in the three-flavor case relevant for the present study, $\SU(6)$ 
spin-flavor effects  provide additional important suppression factors for the final baryon relic density.
The three-flavor benchmark yields for both mass orderings are given
by Eq.~\eqref{eq:cosmo-Qnetwork-yields}, with
$Y_{\mB_\Q}\to Y_{\mB_0}$ in the neutral mixed-flavor branch. 
Under the adopted recombination and cosmological assumptions, in both mass orderings,  
the benchmarks 
at
$\mN=15$, $18$, and $21$ remain below the adopted ten-percent limit, with
constituent-mass estimates
$\Omega_{\mB}h^2\simeq3.4\times10^{-3}$,
$7.6\times10^{-8}$, and $7.6\times10^{-12}$, respectively.  As established
in Sec.~\ref{sec:uv-constraints}, $\mN\leq12$ does not satisfy the combined
baseline requirements; lower $f_a$ or nonstandard depletion can reopen the
$\mN=12$ case.  By $\mN=21$, both the relic abundance and PQ quality are
already comfortably controlled with order-one Yukawas, so larger values
offer no new phenomenological advantage and only increase the pressure on SM
perturbativity.  

The six-panel comparison makes the complementary large-$\mN$
tradeoff explicit: increasing $\mN$ improves the operator-dimension
suppression of PQ breaking and suppresses baryon formation, but worsens the
QCD and hypercharge running.  For $\mN=15$, $18$, and $21$, PQ quality is
achieved before perturbativity is lost, leaving an allowed interval in
$\Lambda_{\rm PQ}$ in both mass orderings.  The $\mN=21$ interval contains
the explicit common-scale point of Eq.~\eqref{eq:common-scale-benchmark}.

Within these dynamical assumptions, the construction supplies a concrete route toward
satisfying simultaneously the quality, minimal-string, BBN, relic, and
perturbativity criteria, thus reconciling axion quality with post-inflation cosmology.
Because PQ breaking occurs after inflation and the string--wall
network is unstable, the initial misalignment angle is statistically fixed
rather than a freely adjustable parameter.  Once the remaining uncertainty
in axion radiation from strings and walls is resolved, the axion abundance
and mass are therefore calculable within the standard cosmological history.

\section*{Acknowledgments} 
We thank Giovanni Villadoro for useful conversations.  
The work of LDL is supported by the Italian Ministry of University and Research (MUR) via the FIS2 Consolidator Grant project FIS-2023-02106 -- QAXION (CUP: I53C25001880001).
LDL and MN are supported by the European Union -- Next Generation EU and
by the Italian Ministry of University and Research (MUR) 
via the PRIN 2022 project n.~2022K4B58X -- AxionOrigins.
The work of EN was supported  by the Estonian Research Council grant PRG1884
and by the INFN ``Iniziativa Specifica'' Theoretical Astroparticle Physics (TAsP).
Partial support from  the Estonian Research Council grants TARISTU24-TK10, TARISTU24-TK3,  CoE grant TK202 “Foundations of the Universe”,
and from the CERN and ESA Science Consortium of Estonia, grants RVTT3 and RVTT7 is also acknowledged.

\appendix

\section{Hypercharge factors and axion-photon coupling}
\label{sec:hypercharge}

\begin{figure*}[t!]
  \centering
  \includegraphics[width=0.8\textwidth]{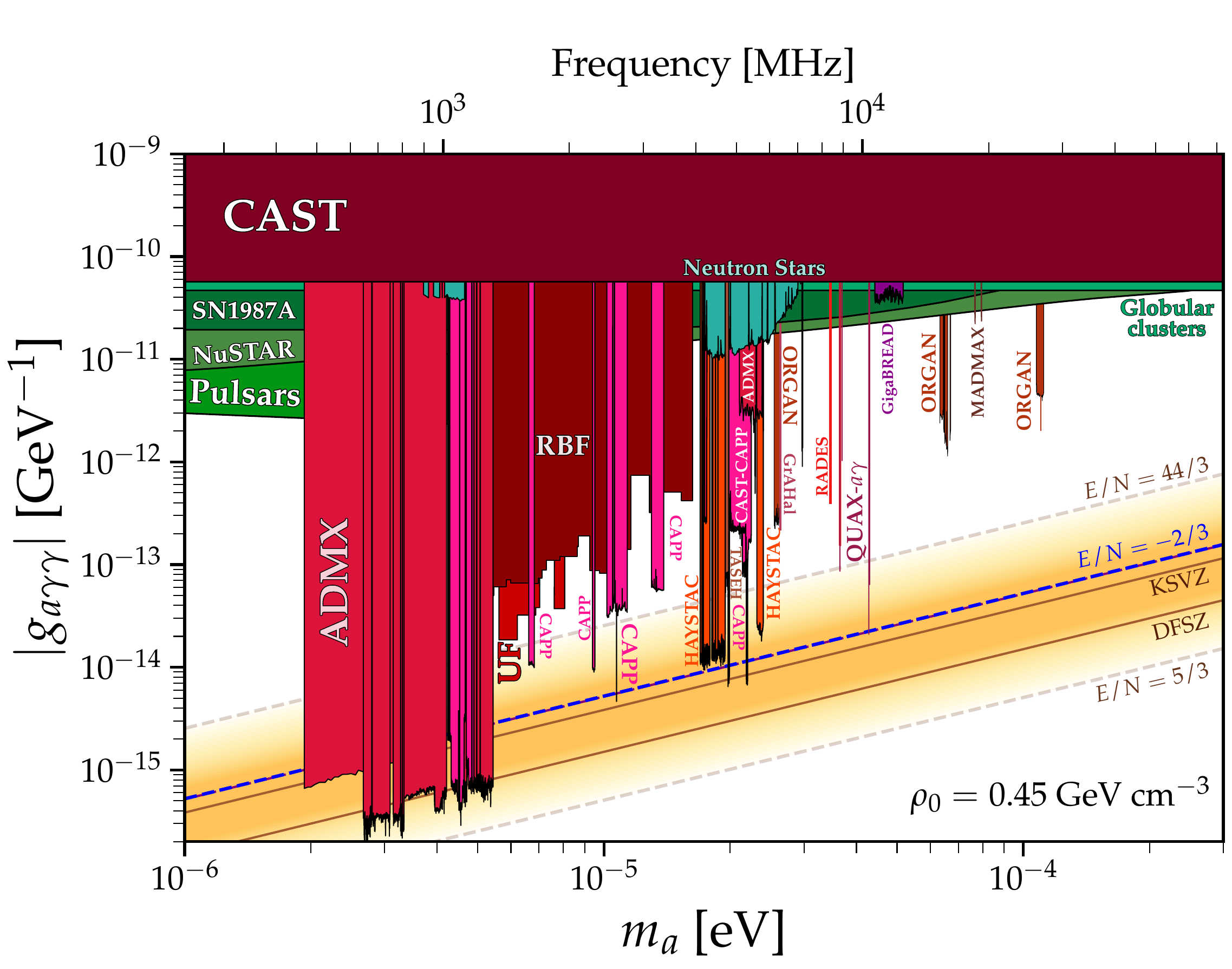}
  \caption{Axion-photon coupling for the minimal assignment
  $E/N=-2/3$ used in the main text (blue line), shown together with the
  canonical yellow axion band $E/N\in[5/3,44/3]$
  \cite{DiLuzio:2016sbl,DiLuzio:2017pfr}.  Figure adapted from
  Ref.~\cite{AxionLimits}.}
  \label{fig:axionphoton}
\end{figure*}

We now discuss in detail the constraints on the hypercharge factors $\Y^a$ in \Table{tab:NewMatterContent}. 
A crucial requirement is the cancellation of the new gauge anomalies involving $\U(1)_Y$. 
SM-related anomalies, such as 
$\U(1)_Y[\mathrm{gravity}]^2$, 
$[\U(1)_Y]^3$ and 
$\U(1)_Y[\SU(3)_C]^2$,  
vanish by construction, since the new representations are 
vector-like under the SM gauge group. 
The only nontrivial condition comes from the $\U(1)_Y[\SU(\N)_{L,R}]^2$ anomaly, which yields
\begin{equation}\label{solNew_YConstraint}
   \A\!\left(\U(1)_Y\left[\SU(\N)_{L,R}\right]^2\right) 
   = \frac{1}{2}\sum_a \Y^a \stackrel{!}{=} 0 \, .
\end{equation}
Other anomaly coefficients, relevant for axion couplings, are
\begin{align}
\label{eq:YPQanom}
&\A\!\left(\U(1)_{\rm PQ}[\SU(3)_C]^2\right) 
   = \frac{\N}{2} \, , \\
&\A\!\left(\U(1)_{\rm PQ}[\U(1)_Y]^2\right) 
   = - \frac{\N}{2}\sum_a \Y_a^2 \, ,
\end{align}
so that the ratio of the electromagnetic to QCD anomalies is 
$E/N = -\sum_a (\Y^a)^2$. 
This ratio determines the axion-photon coupling, 
$\mathcal{L} \supset \frac{1}{4} g_{a\gamma} a F \tilde F$, 
which reads \cite{diCortona:2015ldu,DiLuzio:2020wdo}
\beq 
\label{eq:axionphotoncoupll}
g_{a\gamma} = \frac{\alpha}{2\pi f_a} \( \frac{E}{N} - 1.92(4)  \) \, . 
\eeq
Note that $E/N < 0$ due to the fact that $\Q$ and $\Psi^a$ transform as conjugate representations of $\SU(\N)_{L,R}$ and have opposite PQ charge, 
implying constructive interference with the model-independent contribution in \eq{eq:axionphotoncoupll}.

\begin{table}[h!]
\centering 
\renewcommand{\arraystretch}{1.25}
\begin{tabular}{|c|c|c|c|}
\hline
$\Y^1$ & $\Y^2$ & $\Y^3$ & $E/N$ \\ 
\hline
$-\frac{1}{3}$ & $-\frac{1}{3}$ & $\frac{2}{3}$ & $-\frac{2}{3}$ \\
$-\frac{4}{3}$ & $\frac{2}{3}$ & $\frac{2}{3}$ & $-\frac{8}{3}$ \\
$-\frac{4}{3}$ & $-\frac{1}{3}$ & $\frac{5}{3}$ & $-\frac{14}{3}$ \\
$-\frac{7}{3}$ & $\frac{2}{3}$ & $\frac{5}{3}$ & $-\frac{26}{3}$ \\
$-\frac{4}{3}$ & $-\frac{4}{3}$ & $\frac{8}{3}$ & $-\frac{32}{3}$ \\
$\ldots$ & $\ldots$ & $\ldots$ & $\ldots$ \\
\hline
\end{tabular}
\caption{\label{tab:listY} 
Allowed hypercharge factors $\Y^a$ and the corresponding $E/N$ values.}
\end{table}

We restrict the hypercharges to multiples of $1/3$.  Together with
$\N\equiv 0\pmod{3}$, this ensures that color-singlet baryons have integer electric
charge.  We further require gauge-invariant operators from each of the three
classes in Eq.~\eqref{eq:Ld6}.  These operators connect the $\Psi$ flavors and
transfer the conserved charge between the $\Psi$ and $\Q$ sectors, breaking
the renormalizable $\U(1)^5$ symmetry to
$\U(1)_{\rm PQ}\times\U(1)_{\Q-\Psi}$.  Consequently, only the lightest
carrier of $\U(1)_{\Q-\Psi}$ remains stable on cosmological time scales.
The hypercharge factors that satisfy the above requirement together with 
\eq{solNew_YConstraint} are listed in \Table{tab:listY}, along with the corresponding 
$E/N$ values, ordered by increasing absolute magnitude. 
The first line corresponds to the minimal model adopted in the main
text and is the only phenomenologically viable assignment.  The remaining
non-minimal assignments satisfy the anomaly and operator-selection criteria,
but increase $\sum_a(\Y^a)^2$ and fail phenomenologically because they
generate a low-scale $\U(1)_Y$ Landau pole.

The viable minimal assignment predicts $E/N=-2/3$; its corresponding
axion-photon coupling is shown in the $(m_a,g_\gamma)$ plane in
\fig{fig:axionphoton}.

\section{Pseudo-Goldstone bosons and the axial singlet}
\label{app:pngb}

If a constituent mass lies below $\Lambda_{L+R}$, confinement produces a
chiral multiplet in addition to the ordinary mesons and baryons.  For one
light block of three Dirac flavors, the approximate symmetry breaking is
\begin{equation}
 \U(3)_L\times\U(3)_R\longrightarrow\U(3)_V,
 \label{eq:pngb-chiral-pattern}
\end{equation}
giving an octet of pNGBs and an axial
singlet $\eta'$.  In the chiral regime their characteristic scales are
\begin{equation}
 \begin{aligned}
 m_{\pi_{ij}}^2&\simeq B_0(m_i+m_j)+\Delta_{\rm SM}^{ij},
 &B_0&=\mathcal{O}(\Lambda_{L+R}),\\
 f_\pi&\simeq c_f\frac{\sqrt{\mN}\Lambda_{L+R}}{4\pi},&&
 \end{aligned}
 \label{eq:pngb-masses}
\end{equation}
where $\Delta_{\rm SM}^{ij}$ denotes the QCD or hypercharge contribution,
while
$
 c_f
=\mathcal{O}(1)
$
parametrizes the nonperturbative uncertainty in the large-$\mN$ NDA
normalization of the pNGB decay constant.
For the illustrative confinement choices in
Eq.~\eqref{eq:pngb-benchmark-masses}, only the lighter three-flavor block lies
below confinement.  If both the $\Q$ and $\Psi$ blocks are light, then, after
charge conjugating the $\overline{\mathbf N}$ $\Psi$ fields so that all six
Dirac species are written as fundamentals of the diagonal hidden group, and
neglecting SM gauging, the non-singlet part of the approximate six-flavor
multiplet decomposes as
\begin{equation}
 \mathbf{35}=(\mathbf8,\mathbf1)+(\mathbf1,\mathbf8)
 +(\mathbf3,\overline{\mathbf3})
 +(\overline{\mathbf3},\mathbf3)+(\mathbf1,\mathbf1).
 \label{eq:pngb-six-flavor}
\end{equation}
The mixed modes are not protected by the residual
$\U(1)_{\Q-\Psi}$ and can decay through the $c_i$ operators in
Eq.~\eqref{eq:Ld6}.

\subsection{Anomaly classification}

For light $\Q$, the three QCD-color components play the role of three
chiral flavors from the viewpoint of the confining
$\SU(\mN)_{L+R}$ dynamics.  In the limit $g_s\to0$, they furnish an
approximate $\SU(3)_L\times\SU(3)_R$ symmetry acting on the QCD-color
index; restoring QCD gauges its vector subgroup,
$\SU(3)_V=\SU(3)_c$.  The octet
$\pi_\Q^a\sim\bar\Q t_A^a i\gamma_5\Q$ therefore transforms in the
adjoint of QCD.  Although the axial generator $t_A^a$ and the QCD
generators $T_c^{b,c}$ have different roles, they act on the same
three-dimensional index.  The fermion triangle consequently contains a
single trace over this space.  With
$\operatorname{tr}_{\mathbf 3}(T_c^aT_c^b)=\delta^{ab}/2$ and the same
normalization for $t_A^a$, its anomaly coefficient is
\begin{equation}
 \mathcal A_{\Q}^{a;bc}
 =2\mN\operatorname{tr}_{\mathbf 3}
 \!\left[t_A^a\{T_c^b,T_c^c\}\right]
 =\mN d^{abc}.
 \label{eq:pngb-Q-anomaly}
\end{equation}
Since
$\sum_{bc}d^{abc}d^{a'bc}=(5/3)\delta^{aa'}$, every member of the
octet has a two-gluon decay.  The axial singlet $\eta'_\Q$ also decays
through the QCD anomaly.  Thus no light-$\Q$ pNGB is stable.

For light $\Psi$, the hypercharge matrix is
\begin{equation}
 \mathsf Y=\operatorname{diag}(-1/3,-1/3,2/3),
\end{equation}
and the two-hypercharge-boson anomaly of a neutral pNGB is
$\mathcal A_A=2\mN\operatorname{tr}(t^A\mathsf Y^2)$.  Using
$t^3=\operatorname{diag}(1,-1,0)/2$,
$t^8=\operatorname{diag}(1,1,-2)/(2\sqrt3)$, and
$t^0=\mathbf1_3/\sqrt6$ gives
\begin{equation}
 \mathcal A_3=0,
 \qquad
 \mathcal A_8=-\frac{2\mN}{3\sqrt3},
 \qquad
 \mathcal A_0=\frac{4\mN}{3\sqrt6}.
 \label{eq:pngb-Psi-traces}
\end{equation}
Consequently $\pi_8^0$ and $\eta'_\Psi$ decay directly to two
hypercharge gauge bosons, whereas $\pi_3^0$ and the off-diagonal modes have
no such anomaly.
We normalize the decay constant by
\begin{equation*}
 \langle0|J_{5,A}^\mu|\pi^B(p)\rangle
 =i f_\pi p^\mu\delta^{AB}.
\end{equation*}
With this convention, anomaly matching gives
\begin{equation*}
 \mathcal L_{\rm WZW}\supset
 \frac{\alpha_Y}{4\pi f_\pi}\,
 \mathcal A_A\pi^A B_{\mu\nu}\widetilde B^{\mu\nu}.
\end{equation*}
In particular,
\begin{equation}
 \Gamma(\pi_8^0\to BB)
 =\frac{\mN^2\alpha_Y^2m_8^3}{432\pi^3f_\pi^2}.
 \label{eq:pngb-pi8-width}
\end{equation}
Below electroweak breaking the same interaction gives the corresponding
$\gamma\gamma$, $\gamma Z$, and $ZZ$ channels.  The complete disposition of
the light-$\Psi$ octet is summarized in Table~\ref{tab:pngb-decays}.

\begin{table*}[t]
\centering
\begin{ruledtabular}
\begin{tabular}{lcl}
Mode & $BB$ anomaly & Leading decay path \\
\hline
$\pi_{12},\pi_{21}$
 & no & $a_{12}$ flavor-transfer current \\
$\pi_3^0$
 & no & $\pi_3$--$\pi_8$ mixing from
 $m_{\Psi^1}\ne m_{\Psi^2}$, or explicit flavor transfer \\
$\pi_8^0$
 & yes & prompt anomaly decay \\
$\pi_{i3},\pi_{3i}$ ($i=1,2$)
 & no & $b_i$ current; these are electrically charged \\
$\eta'_\Psi$
 & yes & prompt anomaly decay \\
\end{tabular}
\end{ruledtabular}
\caption{\label{tab:pngb-decays}
Decay classification of the three-flavor light-$\Psi$ chiral multiplet.}
\end{table*}

The mass nondegeneracy already allowed in the Lagrangian induces
$\pi_3$--$\pi_8$ mixing, parametrically giving
\begin{equation}
 \Gamma_{\pi_3}\simeq\theta_{38}^2\Gamma_{\pi_8}.
 \label{eq:pngb-pi3-mixing}
\end{equation}

\subsection{Dimension-six decays and the BBN check}

The anomaly-free modes can decay through the PQ-preserving operators
suppressed by $\Lambda_6$ when the corresponding channel is open.  In the formulas below,
$X\in\{a_{ij},b_i,c_i\}$ labels the relevant operator and its effective
scale defined below Eq.~\eqref{eq:Ld6}.  For a current--current
Lorentz structure, matching the hidden current onto a pNGB gives the
indicative, helicity-suppressed two-body direct-decay rate
\begin{equation}
 \Gamma_{\rm dir}(\pi\to f\bar f')
 \sim\frac{N_c^f}{16\pi}
 \frac{f_\pi^2m_\pi(m_f^2+m_{f'}^2)}
 {(\Lambda_{6,X}^{\rm eff})^4},
\label{eq:pngb-direct-width}
\end{equation}
where $N_c^f$ is the SM color multiplicity of the final state.  Scalar or
tensor operators can have different chiral matching and must be treated
separately.
In the
present benchmarks the top channel is kinematically open whenever it is
selected by the operator.  It gives the fastest available direct channel and
is used here as an illustrative favorable estimate.  A downward transition
between two hidden hadrons instead has the
same phase-space structure as the constituent cascade in
Eq.~\eqref{eq:spectrum-decay-rate},
\begin{equation}
 \Gamma_{i\to f}\sim
 \frac{g_H}{8(192\pi^3)}
 \frac{(\Delta E_{if})^5}{(\Lambda_{6,X}^{\rm eff})^4}.
 \label{eq:pngb-cascade-width}
\end{equation}
The slowest allowed link controls the BBN constraint.  From this
normalization,
\begin{equation}
 \Delta E_{if}\gtrsim9.2\times10^{10}\,\mathrm{GeV}
 \left(\frac{\Lambda_{6,X}^{\rm eff}}{M_{\rm Pl}}\right)^{4/5}
 \left(\frac{100}{g_H}\right)^{1/5},
 \label{eq:pngb-cascade-BBN}
\end{equation}
which becomes $\Delta E_{if}\gtrsim2.0\times10^5\,\mathrm{GeV}$ for
$\Lambda_{6,X}^{\rm eff}=10^{12}\,\mathrm{GeV}$ and $g_H=100$.

As an illustration, the confinement benchmarks below
Eq.~\eqref{eq:cosmo-neutral-confinement-floor} give the NDA estimates
$m_\pi\simeq\sqrt{m_{\rm light}\Lambda_{L+R}}$ and
$f_\pi\simeq\sqrt{\mN}\Lambda_{L+R}/(4\pi)$ (for $c_f=1$):
\begin{equation}
\begin{array}{@{}c|ccc@{}}
\mN&\Lambda_{L+R}&m_\pi&f_\pi\\ \hline
15&10^8\,\mathrm{GeV}&7.4\times10^7\,\mathrm{GeV}&3.1\times10^7\,\mathrm{GeV}\\
18&10^9\,\mathrm{GeV}&7.7\times10^8\,\mathrm{GeV}&3.4\times10^8\,\mathrm{GeV}\\
21&2\times10^{10}\,\mathrm{GeV}&
8.0\times10^{9}\,\mathrm{GeV}&
7.3\times10^{9}\,\mathrm{GeV}
\end{array}
\label{eq:pngb-benchmark-masses}
\end{equation}
The $\mN=15$ and $18$ points lie on the chiral side of the spectrum,
but not parametrically deep inside it:
$m_{\rm light}/\Lambda_{L+R}\simeq0.55$ and $0.60$, respectively.  The
common-scale $\mN=21$ point is better controlled, with
$m_{\rm light}/\Lambda_{L+R}\simeq0.16$.  The displayed pNGB masses and decay
constants should therefore be regarded as NDA estimates with order-one
uncertainty.  A modestly larger confinement scale improves the chiral
expansion and makes the decay rates faster.
For an open top channel and
$\Lambda_{6,X}^{\rm eff}=10^{12}\,\mathrm{GeV}$,
Eq.~\eqref{eq:pngb-direct-width} gives the representative lifetimes
\begin{equation}
 \begin{aligned}
 \tau_{\rm dir}&\sim5.2\times10^{-3}\,\mathrm{s}
 &&(\mN=15),\\
 \tau_{\rm dir}&\sim4.2\times10^{-6}\,\mathrm{s}
 &&(\mN=18).
 \end{aligned}
 \label{eq:pngb-direct-benchmark-lifetimes}
\end{equation}
For the $\mN=21$ common-scale point,
$\Lambda_{6,X}^{\rm eff}=10^{14}\,\mathrm{GeV}$ and the open top channel give
instead
\begin{equation}
 \tau_{\rm dir}\sim8.4\times10^{-2}\,\mathrm{s},
 \qquad (\mN=21),
 \label{eq:pngb-direct-N21-lifetime}
\end{equation}
again safely before BBN.
The anomalous modes are much faster.  These numbers retain order-one
uncertainties from $c_f$, chiral matching, and hadronic form factors.  They
show that, provided every relevant anomaly-free mode has an
order-one coupling to an open third-generation channel, chirality suppression
need not violate the BBN lifetime bound.

\subsection{The large-$\mN$ axial singlet and baryon production}

For three active flavors, the singlet mass is schematically
\begin{equation}
 m_{\eta'}^2\simeq
 \mathcal{O}\!\left(\frac{B_0\operatorname{Tr}M}{n_f}\right)
 +\frac{6\chi_{\rm YM}}{f_{\eta'}^2},
\label{eq:pngb-WV}
\end{equation}
where $M$ is the active-flavor mass matrix. 
The
second term, where  $\chi_{\rm YM}$  denotes  the
pure Yang--Mills topological susceptibility, is the   Witten--Veneziano contribution, that  scales as
$\mathcal{O}(\Lambda_{L+R}^2/\mN)$~\cite{Witten:1979vv,Veneziano:1979ec}.
 In our scenario  the $\eta'$  interacts with the SM gauge fields,  
 since the same constituent
fermions that form the bound state
carry QCD color or hypercharge.  Anomaly matching gives:
\begin{align}
 \Gamma(\eta'_\Q\to gg)
 &=\frac{\mN^2\alpha_s^2m_{\eta'}^3}
 {48\pi^3f_{\eta'}^2},\\
 \Gamma(\eta'_\Psi\to BB)
 &=\frac{\mN^2\alpha_Y^2m_{\eta'}^3}
 {216\pi^3f_{\eta'}^2}, 
 \label{eq:pngb-eta-widths}
\end{align}
where  $gg$ denotes a pair of gluons 
and $BB$ a pair of  $\U(1)_Y$ vector bosons.
Using the same large-$\mN$ NDA normalization as in
Eq.~\eqref{eq:pngb-masses},
$f_{\eta'}\simeq c_\eta\sqrt{\mN}\Lambda_{L+R}/(4\pi)$, with
$c_\eta=\mathcal{O}(1)$.  The factors of $\mN$ in the anomaly then compensate the
large-$\mN$ decrease of $m_{\eta'}$.  In the anomaly-dominated
large-$\mN$ limit,
$\Gamma_{\eta'}/m_{\eta'}=\mathcal{O}(\alpha_{\rm SM}^2/\pi)$, up to group-theory
and NDA coefficients.  The explicit mass contribution only increases this
ratio.  The $\eta'$ is therefore short lived 
and does not contribute to the 
DM energy density balance.

Note that, since $m_{\mB}/m_{\eta'}=\mathcal{O}(\mN^{3/2})$, for $\mN=15,18,21$ the process $\eta'\eta'\to\mB\bar\mB$ is strongly suppressed by the Boltzmann factor, rendering this baryon production channel subleading. We have therefore neglected its contribution to the baryon relic density.

\section{Hadronization model}
\label{app:confinement-hadronization}

The estimates for the baryon yield  in Sec.~\ref{sec:cosmo-confinement} are 
obtained by integration of the   network of Boltzmann equations for the evolution 
of the cluster abundances  developed in Ref.~\cite{inprogress}.  
The  mechanism adopted for modelling  baryon assembly is inspired by 
the QCD Resonance Recombination Model for meson formation, 
\cite{Ravagli:2007xx,Ravagli:2008rt}, and by its extension to  three-body 
cluster formation developed in Ref.~\cite{He:2019vgs}.  

Denoting the fundamental constituent by $\chi=\Q,\Psi^a$, the NN 
steps through the  tower of  $\SU(\mN)$ antisymmetric clusters  are
\begin{align}
 \chi+[\chi^p]_A &\longrightarrow [\chi^{p+1}]_A\,,
 \label{eq:cosmo-cluster-growth}\\
 \bar\chi+[\chi^p]_A &\longrightarrow
 [\chi^{p-1}]_A+[\chi\bar\chi]\,, 
 \label{eq:cosmo-cluster-destruction}
\end{align}
where a cluster of $p$ constituents in the $p$-index
antisymmetric representation is denoted as $[\chi^p]_A$. 
 The tower connects the
fundamental constituent at $p=1$ to the color-singlet baryon at $p=\mN$.
We restrict the orbital wave function to its symmetric $L=0$ ground state.
Fermi statistics then determine the associated  spin--flavor configuration for $n_f=3$.
The nearest-neighbor (NN) network retains only the elementary
single-constituent growth and destruction steps in
Eqs.~\eqref{eq:cosmo-cluster-growth} and
\eqref{eq:cosmo-cluster-destruction}.  
The  BNN  network additionally
includes fusion of any two  quark and two antiquark clusters,  whose total constituent number
satisfies $p+q\leq\mN$, as well as   all quark--antiquark  clusters rearrangements, e.g.  
$[\chi^p] + [\bar \chi^q]  \to [\chi^{p-q}] + [\chi\bar\chi]^q  \ (p\geq q) $.
Note that the BNN network does not include mixed-symmetry final-state clusters. 
However,  for these configurations the potential is either repulsive or weakly attractive,
and thus the related numerical  effects are negligible. 
The network  includes  the  $\SU(6)$ spin--flavor multiplicities corresponding to each initial and final state 
configuration, whose effect is instead numerically relevant.

\begin{table}[t!]
\centering
\begin{tabular}{c|cc}
 $\mN$
 & $y_{\mB,{\rm NN}}$ 
 & $y_{\mB,{\rm BNN}}$\\ \hline
\  \;  $3$ \ 
 & \quad \ $3.5 \times10^{-2\  }$\  \quad & \quad  $3.0\times10^{-2\ }$\\
 $12$
 & \quad $6.7 \times10^{-15}$ & \quad  $7.8 \times10^{-14}$\\
 $15$
 & \quad  $1.7 \times10^{-20}$ & \quad  $3.4\times10^{-19}$\\
 $18$
 & \quad  $2.0\times10^{-26}$ & \quad  $5.6\times10^{-25}$\\
 $21$
 & \quad $1.2\times10^{-32}$ & \quad $4.5\times10^{-31}$ \\
\end{tabular}
\caption{\label{tab:cosmo-network-yields} Baryon-sector  pseudo-yields at the common  stopping time 
$u=200$ for $n_f=3$.  The meaning of  nearest-neighbor (NN) and  
beyond  nearest-neighbor  (BNN) networks is explained in the text.
The charge-symmetric initial conditions $y_1(0)=\bar y_1(0)=1$
 imply  $y_{\bar{\mB}}=y_{\mB}$ for  the antibaryon pseudo-yield.}
\end{table}

The Casimir coefficients for
the NN  growth and destruction reactions are
\begin{align}
 \mathcal C_{p,p+1}&=-\frac{\mN+1}{2\mN}\,p\,,
 &
 \mathcal C_{p,p-1}&=-\frac{\mN+1}{2\mN}\,(\mN-p)\,,
 \label{eq:cosmo-Casimir}
\end{align}
while direct meson formation has
$\mathcal C_{1,0}=-(\mN^2-1)/(2\mN)$. 

 For $p\ll\mN$, cluster growth is
therefore controlled by an $\mathcal{O}(p)$ potential, whereas direct meson formation
and cluster destruction are controlled by an $\mathcal{O}(\mN)$ potential. This
hierarchy is the Casimir bottleneck unveiled in Ref.~\cite{casimirletter}. 

In the network of reactions, all channels are normalized to the fundamental 
coefficient for meson formation $|\mathcal{C}_{1,0}|= C_F$~\cite{casimirletter},
\begin{equation}
 \widehat{\mathcal C}_{ab}\equiv
 \frac{\mathcal C_{ab}}{C_F},\qquad
 C_F=\frac{\mN^2-1}{2\mN},
 \label{eq:cosmo-Casimir-normalization}
\end{equation}
and the kinetic ansatz takes the rates to be proportional to
$|\widehat{\mathcal C}_{ab}|^2$.  The comparison uses the dimensionless
time variable  $u=\Lambda_{L+R}\tau$ at the common stopping point for the integration
$u=200$, and the  initial condition $y_1(0)=\bar y_1(0)=1$.  The resulting
baryon-sector network pseudo-yields evaluated with  the  appropriate 
 $\SU(6)$ spin--flavor multiplicities,  corresponding  to three
degenerate flavors,  are collected in Table~\ref{tab:cosmo-network-yields}.
The results for the NN approximation, given  in the first column, can be   confronted 
with the corresponding results of the BNN network. We see that even for the largest 
values of $\mN$ the NN approximation produces acceptable results, underestimating  the 
baryon yield  just by factors of $\mathcal{O}(10)$. 

More precisely, the additional cluster reactions enhance the three-flavor
pseudo-yield by factors $12$, $19$, $28$, and $39$ at $\mN=12$, $15$, $18$,
and $21$, respectively, but do not remove its strong large-$\mN$ suppression.
The $\SU(6)$ spin--flavor multiplicities   have a   more important numerical 
impact.  If flavor multiplicities are omitted,    
 the BNN  antisymmetric-cluster results would be 
 larger by factors of
 $6.9\times10^2$, $9.1\times10^3$, $1.3\times10^5$, and
$2.1\times10^6$ for $\mN=12$, $15$, $18$,
and $21$, respectively.

Note that  the entries in Table~\ref{tab:cosmo-network-yields} are
network pseudo-yields at a fixed  value of the dimensionless time variable $u=200$, 
not asymptotic hadronic relic yields.  The full network of Casimir-controlled reactions  
cannot account for complete hadronization of free color charges, 
since once the densities have dropped below a certain value, all cluster-(anti)cluster reactions quench.  
More precisely, we find that at $u=200$ the residual pseudo-yield of unbound colored clusters remains of order $y_{\rm col}=\mathcal{O}(10^{-3})$ for all values of $\mN$. This is because the interaction between unscreened color sources approximately obeys Casimir scaling at short and intermediate distances, for both the Coulombic and linear components of the potential. However, as discussed in Sec.~\ref{sec:string-breaking}, at low densities, corresponding to larger distances, string-breaking effects, which are determined by the $\mN$-ality of the source configurations, become the dominant mechanism driving the complete confinement of all naked color charges.
As argued in \cite{casimirletter,inprogress}, at large $\mN$ meson formation through string breaking is strongly favored over 
baryon formation. Therefore, although the cluster network does not completely determine the asymptotic baryon 
pseudo-yields after completion of the confining phase transition, and the uncertainty associated with fragmentation 
must be kept separate from the robust Casimir suppression, string breaking is not expected to significantly 
modify the $u=200$ baryon pseudo-yield values.

\section{Two-loop beta functions and Landau poles}
\label{sec:two-loop-running}

We here provide the beta functions of the 
SM gauge couplings in the presence of the new multiplets 
$\Q \oplus \Psi^a$ and study the emergence
of the associated Landau poles.
The RG equations for the SM gauge couplings, at
two-loop order \cite{Machacek:1983tz}, are
\begin{equation}
\label{alpha2loops}
\mu \frac{{\rm d}}{{\rm d}\mu}\alpha^{-1}_{i}=-\frac{a_{i}}{2\pi}-\frac{b_{ij}}{8\pi^2}\alpha_{j} \, ,
\end{equation}
where $\alpha_i = g_i^2 / (4\pi)$ and 
the beta coefficients are given by
(no summation over $i$) 
\begin{widetext}
\begin{align}
\label{oneloopbf}
a_{i}&=- \frac{11}{3} C_2(G_i) + \frac{4}{3} \sum_F \kappa S_2(F_i) + \frac{1}{3} \sum_S \eta S_2(S_i)\,, \\
\label{twoloopbf}
b_{ij}&= 
\left[- \frac{34}{3} \left( C_2(G_i) \right)^2 
+  \sum_F \left( 4 C_2(F_i) + \frac{20}{3} C_2(G_i) \right) \kappa S_2(F_i) \right.  \\
& \left.  + \sum_S \left( 4 C_2(S_i) + \frac{2}{3} C_2(G_i) \right) \eta S_2(S_i) \right]\delta_{ij} 
+ 4 \Big[  \sum_F \kappa C_2(F_j) S_2(F_i) + \sum_S \eta C_2(S_j) S_2(S_i)  \Big] \, . \nonumber
\end{align}
\end{widetext}
Here, $G_i$ denotes the $i$-th gauge factor, $S_{2}$ and $C_{2}$ are the index (including multiplicity 
factors) and the quadratic Casimir of a given (fermionic ($F$) or scalar ($S$)) 
irreducible representation; $\kappa=1,\frac{1}{2}$ for Dirac and Weyl fermions and
$\eta=1, \frac{1}{2}$ for complex and real scalar fields, respectively. 
The two-loop beta-functions retain the SM gauge-sector
contributions induced by the new matter but neglect mixed terms involving
the hidden gauge couplings, as well as Yukawa contributions.
The coefficient vectors and matrices below are displayed in the
ordered gauge basis
$(\SU(3)_C,\SU(2)_L,\U(1)_1)$, where $\U(1)_1$ uses GUT normalization.
The Yukawa contribution in the two-loop beta function is neglected.
The running involves four effective theories.  The intermediate
theory is SM+$\Q$ when $m_\Q<m_\Psi$ and SM+$\Psi^a$ for the opposite
ordering:
\begin{enumerate}

\item SM 
\begin{align}
a^{\rm SM} &= \left(-7,-\frac{19}{6},\frac{41}{10}\right) \, , \\
b^{\rm SM} &=
\renewcommand{\arraystretch}{1.25}
\left(
\begin{array}{ccc}
 -26 & \frac{9}{2} & \frac{11}{10} \\
 12 & \frac{35}{6} & \frac{9}{10} \\
 \frac{44}{5} & \frac{27}{10} & \frac{199}{50} \\
\end{array}
\right)
\, , 
\end{align}

\item SM + $\Q$
\begin{align}
a^{\rm SM + \Q} &= \left( -7 + \frac{2 \, \N}{3},-\frac{19}{6},\frac{41}{10}\right) \, , \\ 
b^{\rm SM + \Q} &=
\left(
\begin{array}{ccc}
 -26 + \frac{38 \, \N}{3} & \frac{9}{2} & \frac{11}{10}   \phantom{\Big|} \\
 12 & \frac{35}{6} & \frac{9}{10}  \phantom{\Big|} \\
 \frac{44}{5} & \frac{27}{10} & \frac{199}{50} \\
\end{array}
\right)
\, , 
\end{align}

\item SM + $\Psi^a$
\begin{align}
a^{\rm SM+\Psi} &=
\left(-7,-\frac{19}{6},
\frac{41}{10}+\frac{4\N}{5}\sum_a(\Y^a)^2\right),\\
b^{\rm SM+\Psi} &=
\left(
\begin{array}{ccc}
-26 & \frac{9}{2} & \frac{11}{10}   \phantom{\Big|}
\\
12 & \frac{35}{6} & \frac{9}{10}  \phantom{\Big|} \\
\frac{44}{5} & \frac{27}{10} &
\frac{199}{50}+\frac{36\N}{25}\sum_a(\Y^a)^4
\end{array}
\right).
\end{align}

\item SM + $\Q$ + $\Psi^a$
\begin{align}
& a^{\rm SM + \Q + \Psi^a} = \nonumber \\
& \left( -7 + \frac{2 \, \N}{3},-\frac{19}{6},\frac{41}{10} + \frac{4 \, \N}{5} \sum_a (\Y^a)^2 \right)
\, , \\
& b^{\rm SM + \Q + \Psi^a} = \nonumber \\ 
& 
\renewcommand{\arraystretch}{1.25}
\left(
\begin{array}{ccc}
-26 + \frac{38\, \N}{3} & \frac{9}{2} & \frac{11}{10} \\
 12 & \frac{35}{6} & \frac{9}{10} \\
 \frac{44}{5} & \frac{27}{10} & \frac{199}{50} + \frac{36 \, \N}{25} \sum_a (\Y^a)^4 \\
\end{array}
\right) \, ,
\end{align}
\end{enumerate}
where we employed the 
GUT normalization for the abelian factor, 
and we use the values 
$\alpha_1 (m_Z) = 0.016923$,
$\alpha_2 (m_Z) = 0.03374$,
and $\alpha_3 (m_Z) = 0.1173$ for the onset of the RG running \cite{Mihaila:2012pz}. 
Thus the entries of the displayed vectors are ordered as
$(\alpha_3,\alpha_2,\alpha_1)$, whereas the subscripts on the input couplings
retain their conventional SM meaning.  The unnormalized coupling used in the
main-text one-loop formula is $\alpha_Y=3\alpha_1/5$.

It is worth stressing that two-loop running effects are particularly important for the 
strong coupling. The one-loop beta-function coefficient 
$a^{\rm SM+\Q} = -7 + \tfrac{2}{3}\,\N$ 
would vanish at $\N=21/2$ and is consequently accidentally small near the integer values of interest, 
which can drastically affect the scale at which the coupling loses 
perturbativity. The diagonal two-loop coefficient
$(b^{\rm SM+\Q})_{11}=-26+38\N/3$ has no corresponding cancellation in the
range of interest. For the displayed $\Q$-light benchmarks,
hypercharge is the first two-loop pole at $\N=15$, QCD becomes the first pole
at $\N=18$, and hypercharge is again first at $\N=21$ because the heavier
$\Psi$ threshold still carries the enlarged multiplicity.  Hypercharge is
first in all three $\Psi$-light panels, as shown in
Fig.~\ref{fig:paramspace}.  This pattern of limiting couplings is not visible
from the one-loop cancellation alone.
Including two-loop terms is therefore essential for determining the highest
scale to which the minimal field content remains perturbative and for
comparing that scale with the PQ-breaking scale $\Lambda_{\rm PQ}$.  The same
issue is generic in large-multiplicity solutions of the PQ-quality problem
whenever a one-loop coefficient is accidentally small.

 \vfill

\bibliography{bibliography_full}

@article{casimirletter,
    author = "Di Luzio, Luca and Di Valeriano, Samuele and Nardi, Enrico",
    title = "{A Casimir bottleneck in primordial large-$\mathcal{N}$ baryon formation}",
    eprint = "2609.xxxxx",
    archivePrefix = "arXiv",
    primaryClass = "hep-ph",
    year = "2026",
    doi = "",
    journal = "to appear,",
    volume = "",
    number = "",
    pages = "",
}

@unpublished{inprogress,
    author = "Di Luzio, Luca and Di Valeriano, Samuele and Nardi, Enrico",
    title = "{$SU(\mathcal{N})$ baryon formation in the early Universe   
    and dark matter}",
    note = "to appear",
    year = "2026"
}

@article{Choi:2003wr,
    author = "Choi, Ki-woon",
    title = "{A QCD axion from higher dimensional gauge field}",
    eprint = "hep-ph/0308024",
    archivePrefix = "arXiv",
    reportNumber = "KAIST-TH-2003-07",
    doi = "10.1103/PhysRevLett.92.101602",
    journal = "Phys. Rev. Lett.",
    volume = "92",
    pages = "101602",
    year = "2004"
}

@article{Craig:2024dnl,
    author = "Craig, Nathaniel and Kongsore, Marius",
    title = "{High-quality axions from higher-form symmetries in extra dimensions}",
    eprint = "2408.10295",
    archivePrefix = "arXiv",
    primaryClass = "hep-ph",
    doi = "10.1103/PhysRevD.111.015047",
    journal = "Phys. Rev. D",
    volume = "111",
    number = "1",
    pages = "015047",
    year = "2025"
}

@misc{Choi:2026kxu,
    author = "Choi, Kiwoon and Lee, Chang Hyeon and Shin, Chang Sub",
    title = "{Axion Quality in Warped Extra-Dimension}",
    eprint = "2604.08700",
    archivePrefix = "arXiv",
    primaryClass = "hep-ph",
    month = "4",
    year = "2026"
}

@misc{FernandezNavarro:2026bed,
    author = "Fern{\'a}ndez Navarro, Mario",
    title = "{The Minimal High-Quality QCD Axion}",
    eprint = "2607.08814",
    archivePrefix = "arXiv",
    primaryClass = "hep-ph",
    month = "7",
    year = "2026"
}

@article{Juknevich:2009ji,
    author = "Juknevich, Jose E. and Melnikov, Dmitry and Strassler, Matthew J.",
    title = "{A Pure-Glue Hidden Valley I. States and Decays}",
    eprint = "0903.0883",
    archivePrefix = "arXiv",
    primaryClass = "hep-ph",
    reportNumber = "RUNHETC-2008-18, TAUP-2890-08, ITEP-TH-45-08",
    doi = "10.1088/1126-6708/2009/07/055",
    journal = "JHEP",
    volume = "07",
    pages = "055",
    year = "2009"
}

@article{Lazarides:1982tw,
    author = "Lazarides, George and Shafi, Q.",
    title = "{Axion Models with No Domain Wall Problem}",
    reportNumber = "RU82-B-27",
    doi = "10.1016/0370-2693(82)90506-8",
    journal = "Phys. Lett. B",
    volume = "115",
    pages = "21--25",
    year = "1982"
}

@article{Saikawa:2024bta,
    author = "Saikawa, Ken'ichi and Redondo, Javier and Vaquero, Alejandro and Kaltschmidt, Mathieu",
    title = "{Spectrum of global string networks and the axion dark matter mass}",
    eprint = "2401.17253",
    archivePrefix = "arXiv",
    primaryClass = "hep-ph",
    reportNumber = "KANAZAWA-24-02, MPP-2024-18",
    doi = "10.1088/1475-7516/2024/10/043",
    journal = "JCAP",
    volume = "10",
    pages = "043",
    year = "2024"
}

@article{Witten:1979vv,
    author = "Witten, Edward",
    title = "{Current Algebra Theorems for the U(1) Goldstone Boson}",
    reportNumber = "HUTP-79/A014",
    doi = "10.1016/0550-3213(79)90031-2",
    journal = "Nucl. Phys. B",
    volume = "156",
    pages = "269--283",
    year = "1979"
}

@article{Veneziano:1979ec,
    author = "Veneziano, G.",
    title = "{U(1) Without Instantons}",
    reportNumber = "CERN-TH-2651",
    doi = "10.1016/0550-3213(79)90332-8",
    journal = "Nucl. Phys. B",
    volume = "159",
    pages = "213--224",
    year = "1979"
}

@article{DiLuzio:2020qio,
    author = "Di Luzio, Luca",
    title = "{Accidental SO(10) axion from gauged flavour}",
    eprint = "2008.09119",
    archivePrefix = "arXiv",
    primaryClass = "hep-ph",
    reportNumber = "DESY 20-133, DESY-20-133",
    doi = "10.1007/JHEP11(2020)074",
    journal = "JHEP",
    volume = "11",
    pages = "074",
    year = "2020"
}

@article{Gouttenoire:2023roe,
    author = "Gouttenoire, Yann and Kuflik, Eric and Liu, Di",
    title = "{Heavy baryon dark matter from SU(N) confinement: Bubble wall velocity and boundary effects}",
    eprint = "2311.00029",
    archivePrefix = "arXiv",
    primaryClass = "hep-ph",
    reportNumber = "LAPTH-055/23",
    doi = "10.1103/PhysRevD.109.035002",
    journal = "Phys. Rev. D",
    volume = "109",
    number = "3",
    pages = "035002",
    year = "2024"
}

@article{Asadi:2021yml,
    author = "Asadi, Pouya and Kramer, Eric David and Kuflik, Eric and Ridgway, Gregory W. and Slatyer, Tracy R. and Smirnov, Juri",
    title = "{Accidentally Asymmetric Dark Matter}",
    eprint = "2103.09822",
    archivePrefix = "arXiv",
    primaryClass = "hep-ph",
    reportNumber = "MIT-CTP/5284",
    doi = "10.1103/PhysRevLett.127.211101",
    journal = "Phys. Rev. Lett.",
    volume = "127",
    number = "21",
    pages = "211101",
    year = "2021"
}

@article{Asadi:2021pwo,
    author = "Asadi, Pouya and Kramer, Eric David and Kuflik, Eric and Ridgway, Gregory W. and Slatyer, Tracy R. and Smirnov, Juri",
    title = "{Thermal squeezeout of dark matter}",
    eprint = "2103.09827",
    archivePrefix = "arXiv",
    primaryClass = "hep-ph",
    doi = "10.1103/PhysRevD.104.095013",
    journal = "Phys. Rev. D",
    volume = "104",
    number = "9",
    pages = "095013",
    year = "2021"
}

@article{Ravagli:2007xx,
    author = "Ravagli, L. and Rapp, R.",
    title = "{Quark Coalescence based on a Transport Equation}",
    eprint = "0705.0021",
    archivePrefix = "arXiv",
    primaryClass = "hep-ph",
    doi = "10.1016/j.physletb.2007.07.043",
    journal = "Phys. Lett. B",
    volume = "655",
    pages = "126--131",
    year = "2007"
}

@article{Ravagli:2008rt,
    author = "Ravagli, L. and van Hees, H. and Rapp, R.",
    title = "{Resonance Recombination Model: A Dynamical Framework for Hadronization}",
    eprint = "0806.2055",
    archivePrefix = "arXiv",
    primaryClass = "hep-ph",
    doi = "10.1103/PhysRevC.79.064902",
    journal = "Phys. Rev. C",
    volume = "79",
    pages = "064902",
    year = "2009"
}

@article{He:2019vgs,
    author = "He, Min and Rapp, Ralf",
    title = "{Hadronization and Charm-Hadron Ratios in Heavy-Ion Collisions}",
    eprint = "1905.09216",
    archivePrefix = "arXiv",
    primaryClass = "nucl-th",
    doi = "10.1103/PhysRevLett.124.042301",
    journal = "Phys. Rev. Lett.",
    volume = "124",
    number = "4",
    pages = "042301",
    year = "2020"
}

@article{Gasser:2020mzy,
    author = "Gasser, J. and Leutwyler, H. and Rusetsky, A.",
    title = "{On the mass difference between proton and neutron}",
    eprint = "2003.13612",
    archivePrefix = "arXiv",
    primaryClass = "hep-ph",
    doi = "10.1016/j.physletb.2021.136087",
    journal = "Phys. Lett. B",
    volume = "814",
    pages = "136087",
    year = "2021"
}

@article{DeGrand:2021zjw,
    author = "DeGrand, Thomas",
    title = "{Finite temperature properties of QCD with two flavors and three, four and five colors}",
    eprint = "2102.01150",
    archivePrefix = "arXiv",
    primaryClass = "hep-lat",
    doi = "10.1103/PhysRevD.103.094513",
    journal = "Phys. Rev. D",
    volume = "103",
    number = "9",
    pages = "094513",
    year = "2021"
}

@article{Ayyar:2018ppa,
    author = "Ayyar, Venkitesh and DeGrand, Thomas and Hackett, Daniel C. and Jay, William I. and Neil, Ethan T. and Shamir, Yigal and Svetitsky, Benjamin",
    title = "{Finite-temperature phase structure of SU(4) gauge theory with multiple fermion representations}",
    eprint = "1802.09644",
    archivePrefix = "arXiv",
    primaryClass = "hep-lat",
    doi = "10.1103/PhysRevD.97.114502",
    journal = "Phys. Rev. D",
    volume = "97",
    number = "11",
    pages = "114502",
    year = "2018"
}

@article{DeGrand:2018tzn,
    author = "DeGrand, Thomas and Hackett, Daniel C. and Neil, Ethan T.",
    title = "{Large $N_c$ Thermodynamics with Dynamical Fermions}",
    eprint = "1809.00073",
    archivePrefix = "arXiv",
    primaryClass = "hep-lat",
    doi = "10.22323/1.334.0175",
    journal = "PoS",
    volume = "LATTICE2018",
    pages = "175",
    year = "2018"
}

@article{LatticeStrongDynamics:2020jwi,
    author = "Brower, R. C. and others",
    collaboration = "Lattice Strong Dynamics",
    title = "{Stealth dark matter confinement transition and gravitational waves}",
    eprint = "2006.16429",
    archivePrefix = "arXiv",
    primaryClass = "hep-lat",
    reportNumber = "LLNL-JRNL-811356; RIKEN-iTHEMS-Report-20",
    doi = "10.1103/PhysRevD.103.014505",
    journal = "Phys. Rev. D",
    volume = "103",
    number = "1",
    pages = "014505",
    year = "2021"
}

@article{Lucini:2005vg,
    author = "Lucini, Biagio and Teper, Michael and Wenger, Urs",
    title = "{Properties of the deconfining phase transition in SU(N) gauge theories}",
    eprint = "hep-lat/0502003",
    archivePrefix = "arXiv",
    reportNumber = "DESY-05-021",
    doi = "10.1088/1126-6708/2005/02/033",
    journal = "JHEP",
    volume = "02",
    pages = "033",
    year = "2005"
}

@article{Datta:2009jn,
    author = "Datta, Saumen and Gupta, Sourendu",
    title = "{Scaling and the continuum limit of the finite temperature deconfinement transition in SU$(N_c)$ pure gauge theory}",
    eprint = "0909.5591",
    archivePrefix = "arXiv",
    primaryClass = "hep-lat",
    reportNumber = "TIFR-TH-09-34",
    doi = "10.1103/PhysRevD.80.114504",
    journal = "Phys. Rev. D",
    volume = "80",
    pages = "114504",
    year = "2009"
}

@article{Lucini:2012gg,
    author = "Lucini, Biagio and Panero, Marco",
    title = "{SU(N) gauge theories at large N}",
    eprint = "1210.4997",
    archivePrefix = "arXiv",
    primaryClass = "hep-th",
    reportNumber = "HIP-2012-24-TH, NSF-KITP-12-190",
    doi = "10.1016/j.physrep.2013.01.001",
    journal = "Phys. Rept.",
    volume = "526",
    pages = "93--163",
    year = "2013"
}

@article{Saito:2011fs,
    author = "Saito, H. and Ejiri, S. and Aoki, S. and Hatsuda, T. and Kanaya, K. and Maezawa, Y. and Ohno, H. and Umeda, T.",
    collaboration = "WHOT-QCD",
    title = "{Phase structure of finite temperature QCD in the heavy quark region}",
    eprint = "1106.0974",
    archivePrefix = "arXiv",
    primaryClass = "hep-lat",
    reportNumber = "UTHEP-629, RIKEN-MP-22",
    doi = "10.1103/PhysRevD.85.079902",
    journal = "Phys. Rev. D",
    volume = "84",
    pages = "054502",
    year = "2011",
    note = "[Erratum: Phys.Rev.D 85, 079902 (2012)]"
}

@article{Witten:1984rs,
    author = "Witten, Edward",
    title = "{Cosmic Separation of Phases}",
    reportNumber = "PRINT-84-0400 (IAS,PRINCETON)",
    doi = "10.1103/PhysRevD.30.272",
    journal = "Phys. Rev. D",
    volume = "30",
    pages = "272--285",
    year = "1984"
}

@article{Svetitsky:1982gs,
    author = "Svetitsky, Benjamin and Yaffe, Laurence G.",
    title = "{Critical Behavior at Finite Temperature Confinement Transitions}",
    reportNumber = "CLNS-82/530, NSF-ITP-82-53",
    doi = "10.1016/0550-3213(82)90172-9",
    journal = "Nucl. Phys. B",
    volume = "210",
    pages = "423--447",
    year = "1982"
}

@article{Alexandrou:1998wv,
    author = "Alexandrou, Constantia and Borici, Artan and Feo, Alessandra and de Forcrand, Philippe and Galli, Andrea and Jegerlehner, Fred and Takaishi, Tetsuya",
    title = "{The Deconfinement phase transition in one flavor QCD}",
    eprint = "hep-lat/9811028",
    archivePrefix = "arXiv",
    reportNumber = "UCY-PHY-98-05",
    doi = "10.1103/PhysRevD.60.034504",
    journal = "Phys. Rev. D",
    volume = "60",
    pages = "034504",
    year = "1999"
}

@article{Machacek:1983tz,
    author = "Machacek, Marie E. and Vaughn, Michael T.",
    title = "{Two Loop Renormalization Group Equations in a General Quantum Field Theory. 1. Wave Function Renormalization}",
    reportNumber = "NUB-2590, HUTP-83/A003",
    doi = "10.1016/0550-3213(83)90610-7",
    journal = "Nucl. Phys. B",
    volume = "222",
    pages = "83--103",
    year = "1983"
}

@article{Mihaila:2012pz,
    author = "Mihaila, Luminita N. and Salomon, Jens and Steinhauser, Matthias",
    title = "{Renormalization constants and beta functions for the gauge couplings of the Standard Model to three-loop order}",
    eprint = "1208.3357",
    archivePrefix = "arXiv",
    reportNumber = "SFB-CPP-12-61, TTP12-30",
    doi = "10.1103/PhysRevD.86.096008",
    journal = "Phys. Rev. D",
    volume = "86",
    pages = "096008",
    year = "2012"
}

@article{DiLuzio:2024xnt,
    author = "Di Luzio, Luca and Hoof, Sebastian and Marinissen, Coenraad and Plakkot, Vaisakh",
    title = "{Catalogues of cosmologically self-consistent hadronic QCD axion models}",
    eprint = "2412.17896",
    archivePrefix = "arXiv",
    primaryClass = "hep-ph",
    doi = "10.1088/1475-7516/2025/04/072",
    journal = "JCAP",
    volume = "04",
    pages = "072",
    year = "2025"
}

@misc{AxionLimits,
  author       = {Ciaran O'Hare},
  title        = {cajohare/AxionLimits: AxionLimits},
  month        = jul,
  year         = 2020,
  publisher    = {Zenodo},
  version      = {v1.0},
  doi          = {10.5281/zenodo.3932430},
  howpublished = {\url{https://cajohare.github.io/AxionLimits/}}
}

@article{Cheek:2023fht,
    author = "Cheek, Andrew and Osi{\'n}ski, Jacek K. and Roszkowski, Leszek",
    title = "{Extending preferred axion models via heavy-quark induced early matter domination}",
    eprint = "2310.16087",
    archivePrefix = "arXiv",
    primaryClass = "hep-ph",
    doi = "10.1088/1475-7516/2024/03/061",
    journal = "JCAP",
    volume = "03",
    pages = "061",
    year = "2024"
}

@article{Petrossian-Byrne:2025mto,
    author = "Petrossian-Byrne, Rudin and Villadoro, Giovanni",
    title = "{Open string axiverse}",
    eprint = "2503.16387",
    archivePrefix = "arXiv",
    primaryClass = "hep-ph",
    doi = "10.1007/JHEP07(2025)049",
    journal = "JHEP",
    volume = "07",
    pages = "049",
    year = "2025"
}

@article{Borsanyi:2016ksw,
    author = "Borsanyi, Sz. and others",
    title = "{Calculation of the axion mass based on high-temperature lattice quantum chromodynamics}",
    eprint = "1606.07494",
    archivePrefix = "arXiv",
    primaryClass = "hep-lat",
    reportNumber = "DESY-16-105",
    doi = "10.1038/nature20115",
    journal = "Nature",
    volume = "539",
    number = "7627",
    pages = "69--71",
    year = "2016"
}

@article{Hiramatsu:2012gg,
    author = "Hiramatsu, Takashi and Kawasaki, Masahiro and Saikawa, Ken'ichi and Sekiguchi, Toyokazu",
    title = "{Production of dark matter axions from collapse of string-wall systems}",
    eprint = "1202.5851",
    archivePrefix = "arXiv",
    primaryClass = "hep-ph",
    reportNumber = "ICRR-REPORT-608-2011-25, IPMU12-0025, YITP-12-9",
    doi = "10.1103/PhysRevD.85.105020",
    journal = "Phys. Rev. D",
    volume = "85",
    pages = "105020",
    year = "2012",
    note = "[Erratum: Phys.Rev.D 86, 089902 (2012)]"
}

@article{Klaer:2017ond,
    author = "Klaer, Vincent B. . and Moore, Guy D.",
    title = "{The dark-matter axion mass}",
    eprint = "1708.07521",
    archivePrefix = "arXiv",
    primaryClass = "hep-ph",
    doi = "10.1088/1475-7516/2017/11/049",
    journal = "JCAP",
    volume = "11",
    pages = "049",
    year = "2017"
}

@article{Gorghetto:2018myk,
    author = "Gorghetto, Marco and Hardy, Edward and Villadoro, Giovanni",
    title = "{Axions from Strings: the Attractive Solution}",
    eprint = "1806.04677",
    archivePrefix = "arXiv",
    primaryClass = "hep-ph",
    doi = "10.1007/JHEP07(2018)151",
    journal = "JHEP",
    volume = "07",
    pages = "151",
    year = "2018"
}

@article{Gorghetto:2020qws,
    author = "Gorghetto, Marco and Hardy, Edward and Villadoro, Giovanni",
    title = "{More axions from strings}",
    eprint = "2007.04990",
    archivePrefix = "arXiv",
    primaryClass = "hep-ph",
    doi = "10.21468/SciPostPhys.10.2.050",
    journal = "SciPost Phys.",
    volume = "10",
    number = "2",
    pages = "050",
    year = "2021"
}

@article{Reece:2025thc,
    author = "Reece, Matthew",
    title = "{Extra-dimensional axion expectations}",
    eprint = "2406.08543",
    archivePrefix = "arXiv",
    primaryClass = "hep-ph",
    doi = "10.1007/JHEP07(2025)130",
    journal = "JHEP",
    volume = "07",
    pages = "130",
    year = "2025"
}

@article{Buschmann:2021sdq,
    author = "Buschmann, Malte and Foster, Joshua W. and Hook, Anson and Peterson, Adam and Willcox, Don E. and Zhang, Weiqun and Safdi, Benjamin R.",
    title = "{Dark matter from axion strings with adaptive mesh refinement}",
    eprint = "2108.05368",
    archivePrefix = "arXiv",
    primaryClass = "hep-ph",
    doi = "10.1038/s41467-022-28669-y",
    journal = "Nature Commun.",
    volume = "13",
    number = "1",
    pages = "1049",
    year = "2022"
}

@article{Benabou:2024msj,
    author = "Benabou, Joshua N. and Buschmann, Malte and Foster, Joshua W. and Safdi, Benjamin R.",
    title = "{Axion Mass Prediction from Adaptive Mesh Refinement Cosmological Lattice Simulations}",
    eprint = "2412.08699",
    archivePrefix = "arXiv",
    primaryClass = "hep-ph",
    reportNumber = "FERMILAB-PUB-24-0912-T",
    doi = "10.1103/6v21-d6sj",
    journal = "Phys. Rev. Lett.",
    volume = "134",
    number = "24",
    pages = "241003",
    year = "2025"
}

@article{Randall:1992ut,
    author = "Randall, Lisa",
    title = "{Composite axion models and Planck scale physics}",
    reportNumber = "MIT-CTP-2074",
    doi = "10.1016/0370-2693(92)91928-3",
    journal = "Phys. Lett. B",
    volume = "284",
    pages = "77--80",
    year = "1992"
}

@article{Dobrescu:1996jp,
    author = "Dobrescu, Bogdan A.",
    title = "{The Strong CP problem versus Planck scale physics}",
    eprint = "hep-ph/9609221",
    archivePrefix = "arXiv",
    reportNumber = "BUHEP-96-30",
    doi = "10.1103/PhysRevD.55.5826",
    journal = "Phys. Rev. D",
    volume = "55",
    pages = "5826--5833",
    year = "1997"
}

@article{Babu:2002ic,
    author = "Babu, K. S. and Gogoladze, Ilia and Wang, Kai",
    title = "{Stabilizing the axion by discrete gauge symmetries}",
    eprint = "hep-ph/0212339",
    archivePrefix = "arXiv",
    reportNumber = "OSU-HEP-02-18",
    doi = "10.1016/S0370-2693(03)00411-8",
    journal = "Phys. Lett. B",
    volume = "560",
    pages = "214--222",
    year = "2003"
}

@article{Redi:2016esr,
    author = "Redi, Michele and Sato, Ryosuke",
    title = "{Composite Accidental Axions}",
    eprint = "1602.05427",
    archivePrefix = "arXiv",
    primaryClass = "hep-ph",
    doi = "10.1007/JHEP05(2016)104",
    journal = "JHEP",
    volume = "05",
    pages = "104",
    year = "2016"
}

@article{Fukuda:2017ylt,
    author = "Fukuda, Hajime and Ibe, Masahiro and Suzuki, Motoo and Yanagida, Tsutomu T.",
    title = "{A ''gauged'' $U(1)$ Peccei{\textendash}Quinn symmetry}",
    eprint = "1703.01112",
    archivePrefix = "arXiv",
    primaryClass = "hep-ph",
    reportNumber = "IPMU-17-0040",
    doi = "10.1016/j.physletb.2017.05.071",
    journal = "Phys. Lett. B",
    volume = "771",
    pages = "327--331",
    year = "2017"
}

@article{Bonnefoy:2018ibr,
    author = "Bonnefoy, Quentin and Dudas, Emilian and Pokorski, Stefan",
    title = "{Axions in a highly protected gauge symmetry model}",
    eprint = "1804.01112",
    archivePrefix = "arXiv",
    primaryClass = "hep-ph",
    doi = "10.1140/epjc/s10052-018-6528-z",
    journal = "Eur. Phys. J. C",
    volume = "79",
    number = "1",
    pages = "31",
    year = "2019"
}

@article{Lillard:2018fdt,
    author = "Lillard, Benjamin and Tait, Tim M. P.",
    title = "{A High Quality Composite Axion}",
    eprint = "1811.03089",
    archivePrefix = "arXiv",
    primaryClass = "hep-ph",
    reportNumber = "UCI-HEP-TR-2018-12",
    doi = "10.1007/JHEP11(2018)199",
    journal = "JHEP",
    volume = "11",
    pages = "199",
    year = "2018"
}

@article{Gavela:2018paw,
    author = "Gavela, M. B. and Ibe, M. and Quilez, P. and Yanagida, T. T.",
    title = "{Automatic Peccei{\textendash}Quinn symmetry}",
    eprint = "1812.08174",
    archivePrefix = "arXiv",
    primaryClass = "hep-ph",
    reportNumber = "IPMU18-0205, IFT-UAM/CSIC-18-129, FTUAM-18-29",
    doi = "10.1140/epjc/s10052-019-7046-3",
    journal = "Eur. Phys. J. C",
    volume = "79",
    number = "6",
    pages = "542",
    year = "2019"
}

@article{Lee:2018yak,
    author = "Lee, Hye-Sung and Yin, Wen",
    title = "{Peccei-Quinn symmetry from a hidden gauge group structure}",
    eprint = "1811.04039",
    archivePrefix = "arXiv",
    primaryClass = "hep-ph",
    doi = "10.1103/PhysRevD.99.015041",
    journal = "Phys. Rev. D",
    volume = "99",
    number = "1",
    pages = "015041",
    year = "2019"
}

@article{Ardu:2020qmo,
    author = "Ardu, Marco and Di Luzio, Luca and Landini, Giacomo and Strumia, Alessandro and Teresi, Daniele and Wang, Jin-Wei",
    title = "{Axion quality from the (anti)symmetric of SU($ \mathcal{N} $)}",
    eprint = "2007.12663",
    archivePrefix = "arXiv",
    primaryClass = "hep-ph",
    reportNumber = "DESY 20-124, DESY-20-124",
    doi = "10.1007/JHEP11(2020)090",
    journal = "JHEP",
    volume = "11",
    pages = "090",
    year = "2020"
}

@article{Yin:2020dfn,
    author = "Yin, Wen",
    title = "{Scale and quality of Peccei-Quinn symmetry and weak gravity conjectures}",
    eprint = "2007.13320",
    archivePrefix = "arXiv",
    primaryClass = "hep-ph",
    doi = "10.1007/JHEP10(2020)032",
    journal = "JHEP",
    volume = "10",
    pages = "032",
    year = "2020"
}

@article{Contino:2021ayn,
    author = "Contino, Roberto and Podo, Alessandro and Revello, Filippo",
    title = "{Chiral models of composite axions and accidental Peccei-Quinn symmetry}",
    eprint = "2112.09635",
    archivePrefix = "arXiv",
    primaryClass = "hep-ph",
    doi = "10.1007/JHEP04(2022)180",
    journal = "JHEP",
    volume = "04",
    pages = "180",
    year = "2022"
}

@article{DiLuzio:2025jhv,
    author = "Di Luzio, Luca and Landini, Giacomo and Mescia, Federico and Susi{\v{c}}, Vasja",
    title = "{High-quality Peccei-Quinn symmetry from the interplay of vertical and horizontal gauge symmetries}",
    eprint = "2503.16648",
    archivePrefix = "arXiv",
    primaryClass = "hep-ph",
    doi = "10.1140/epjc/s10052-025-15175-w",
    journal = "Eur. Phys. J. C",
    volume = "86",
    number = "1",
    pages = "5",
    year = "2026"
}

@article{Georgi:1981pu,
    author = "Georgi, Howard M. and Hall, Lawrence J. and Wise, Mark B.",
    title = "{Grand Unified Models With an Automatic {Peccei-Quinn} Symmetry}",
    reportNumber = "HUTP-81/A031",
    doi = "10.1016/0550-3213(81)90433-8",
    journal = "Nucl. Phys. B",
    volume = "192",
    pages = "409--416",
    year = "1981"
}

@article{Dine:1986bg,
    author = "Dine, Michael and Seiberg, Nathan",
    title = "{String Theory and the Strong {CP} Problem}",
    reportNumber = "Print-86-0091 (CITY COLL.,N.Y.), CCNY-HEP-86/2",
    doi = "10.1016/0550-3213(86)90043-X",
    journal = "Nucl. Phys. B",
    volume = "273",
    pages = "109--124",
    year = "1986"
}

@article{Ghigna:1992iv,
    author = "Ghigna, S. and Lusignoli, Maurizio and Roncadelli, M.",
    title = "{Instability of the invisible axion}",
    reportNumber = "ROME-877-1992, FNT-T-92-12",
    doi = "10.1016/0370-2693(92)90019-Z",
    journal = "Phys. Lett. B",
    volume = "283",
    pages = "278--281",
    year = "1992"
}

@article{DiLuzio:2017tjx,
    author = "Di Luzio, Luca and Nardi, Enrico and Ubaldi, Lorenzo",
    title = "{Accidental Peccei-Quinn symmetry protected to arbitrary order}",
    eprint = "1704.01122",
    archivePrefix = "arXiv",
    primaryClass = "hep-ph",
    reportNumber = "IPPP-17-29",
    doi = "10.1103/PhysRevLett.119.011801",
    journal = "Phys. Rev. Lett.",
    volume = "119",
    number = "1",
    pages = "011801",
    year = "2017"
}

@article{Lu:2023ayc,
    author = "Lu, Qianshu and Reece, Matthew and Sun, Zhiquan",
    title = "{The quality/cosmology tension for a post-inflation QCD axion}",
    eprint = "2312.07650",
    archivePrefix = "arXiv",
    primaryClass = "hep-ph",
    reportNumber = "MIT-CTP 5644",
    doi = "10.1007/JHEP07(2024)227",
    journal = "JHEP",
    volume = "07",
    pages = "227",
    year = "2024"
}

@article{DiLuzio:2017pfr,
    author = "Di Luzio, Luca and Mescia, Federico and Nardi, Enrico",
    title = "{Window for preferred axion models}",
    eprint = "1705.05370",
    archivePrefix = "arXiv",
    primaryClass = "hep-ph",
    reportNumber = "IPPP-17-41",
    doi = "10.1103/PhysRevD.96.075003",
    journal = "Phys. Rev. D",
    volume = "96",
    number = "7",
    pages = "075003",
    year = "2017"
}

@article{DiLuzio:2016sbl,
    author = "Di Luzio, Luca and Mescia, Federico and Nardi, Enrico",
    title = "{Redefining the Axion Window}",
    eprint = "1610.07593",
    archivePrefix = "arXiv",
    primaryClass = "hep-ph",
    reportNumber = "IPPP-16-99",
    doi = "10.1103/PhysRevLett.118.031801",
    journal = "Phys. Rev. Lett.",
    volume = "118",
    number = "3",
    pages = "031801",
    year = "2017"
}

@article{Carenza:2019pxu,
    author = "Carenza, Pierluca and Fischer, Tobias and Giannotti, Maurizio and Guo, Gang and Mart\'\i{}nez-Pinedo, Gabriel and Mirizzi, Alessandro",
    title = "{Improved axion emissivity from a supernova via nucleon-nucleon bremsstrahlung}",
    eprint = "1906.11844",
    archivePrefix = "arXiv",
    primaryClass = "hep-ph",
    doi = "10.1088/1475-7516/2019/10/016",
    journal = "JCAP",
    volume = "10",
    number = "10",
    pages = "016",
    year = "2019",
    note = "[Erratum: JCAP 05, E01 (2020)]"
}

@article{diCortona:2015ldu,
    author = "Grilli di Cortona, Giovanni and Hardy, Edward and Pardo Vega, Javier and Villadoro, Giovanni",
    title = "{The QCD axion, precisely}",
    eprint = "1511.02867",
    archivePrefix = "arXiv",
    primaryClass = "hep-ph",
    doi = "10.1007/JHEP01(2016)034",
    journal = "JHEP",
    volume = "01",
    pages = "034",
    year = "2016"
}

@article{Dine:1982ah,
      author         = "Dine, Michael and Fischler, Willy",
      title          = "{The Not So Harmless Axion}",
      journal        = "Phys. Lett.",
      volume         = "120B",
      year           = "1983",
      pages          = "137-141",
      doi            = "10.1016/0370-2693(83)90639-1",
      reportNumber   = "UPR-0201T",
      SLACcitation   = "%%CITATION = PHLTA,120B,137;%%"
}

@article{Abbott:1982af,
      author         = "Abbott, L. F. and Sikivie, P.",
      title          = "{A Cosmological Bound on the Invisible Axion}",
      journal        = "Phys. Lett.",
      volume         = "120B",
      year           = "1983",
      pages          = "133-136",
      doi            = "10.1016/0370-2693(83)90638-X",
      reportNumber   = "PRINT-82-0695 (BRANDEIS)",
      SLACcitation   = "%%CITATION = PHLTA,120B,133;%%"
}

@article{Preskill:1982cy,
      author         = "Preskill, John and Wise, Mark B. and Wilczek, Frank",
      title          = "{Cosmology of the Invisible Axion}",
      journal        = "Phys. Lett.",
      volume         = "120B",
      year           = "1983",
      pages          = "127-132",
      doi            = "10.1016/0370-2693(83)90637-8",
      reportNumber   = "HUTP-82-A048, NSF-ITP-82-103",
      SLACcitation   = "%%CITATION = PHLTA,120B,127;%%"
}

@inproceedings{Adams:2022pbo,
    author = "Adams, C. B. and others",
    title = "{Axion Dark Matter}",
    booktitle = "{Snowmass 2021}",
    eprint = "2203.14923",
    archivePrefix = "arXiv",
    primaryClass = "hep-ex",
    reportNumber = "FERMILAB-CONF-22-996-PPD-T",
    month = "3",
    year = "2022"
}

@article{Sikivie:2020zpn,
    author = "Sikivie, Pierre",
    title = "{Invisible Axion Search Methods}",
    eprint = "2003.02206",
    archivePrefix = "arXiv",
    primaryClass = "hep-ph",
    doi = "10.1103/RevModPhys.93.015004",
    journal = "Rev. Mod. Phys.",
    volume = "93",
    number = "1",
    pages = "015004",
    year = "2021"
}

@article{DiLuzio:2020wdo,
    author = "Di Luzio, Luca and Giannotti, Maurizio and Nardi, Enrico and Visinelli, Luca",
    title = "{The landscape of QCD axion models}",
    eprint = "2003.01100",
    archivePrefix = "arXiv",
    primaryClass = "hep-ph",
    reportNumber = "DESY 20-036, DESY-20-036",
    doi = "10.1016/j.physrep.2020.06.002",
    journal = "Phys. Rept.",
    volume = "870",
    pages = "1--117",
    year = "2020"
}

@article{Irastorza:2018dyq,
      author         = "Irastorza, Igor G. and Redondo, Javier",
      title          = "{New experimental approaches in the search for axion-like
                        particles}",
      journal        = "Prog. Part. Nucl. Phys.",
      volume         = "102",
      year           = "2018",
      pages          = "89-159",
      doi            = "10.1016/j.ppnp.2018.05.003",
      eprint         = "1801.08127",
      archivePrefix  = "arXiv",
      primaryClass   = "hep-ph",
      SLACcitation   = "%%CITATION = ARXIV:1801.08127;%%"
}

@article{Sikivie:1982qv,
      author         = "Sikivie, P.",
      title          = "{Of Axions, Domain Walls and the Early Universe}",
      journal        = "Phys. Rev. Lett.",
      volume         = "48",
      year           = "1982",
      pages          = "1156-1159",
      doi            = "10.1103/PhysRevLett.48.1156",
      reportNumber   = "UFTP-82-3",
      SLACcitation   = "%%CITATION = PRLTA,48,1156;%%"
}

@article{Peccei:1977ur,
      author         = "Peccei, R. D. and Quinn, Helen R.",
      title          = "{Constraints Imposed by CP Conservation in the Presence
                        of Instantons}",
      journal        = "Phys. Rev.",
      volume         = "D16",
      year           = "1977",
      pages          = "1791-1797",
      doi            = "10.1103/PhysRevD.16.1791",
      reportNumber   = "ITP-572-STANFORD",
      SLACcitation   = "%%CITATION = PHRVA,D16,1791;%%"
}

@article{Peccei:1977hh,
      author         = "Peccei, R. D. and Quinn, Helen R.",
      title          = "{CP Conservation in the Presence of Instantons}",
      journal        = "Phys. Rev. Lett.",
      volume         = "38",
      year           = "1977",
      pages          = "1440-1443",
      doi            = "10.1103/PhysRevLett.38.1440",
      reportNumber   = "ITP-568-STANFORD",
      SLACcitation   = "%%CITATION = PRLTA,38,1440;%%"
}

@article{Wilczek:1977pj,
      author         = "Wilczek, Frank",
      title          = "{Problem of Strong p and t Invariance in the Presence of
                        Instantons}",
      journal        = "Phys. Rev. Lett.",
      volume         = "40",
      year           = "1978",
      pages          = "279-282",
      doi            = "10.1103/PhysRevLett.40.279",
      reportNumber   = "Print-77-0939 (COLUMBIA)",
      SLACcitation   = "%%CITATION = PRLTA,40,279;%%"
}

@article{Barr:1992qq,
      author         = "Barr, Stephen M. and Seckel, D.",
      title          = "{Planck scale corrections to axion models}",
      journal        = "Phys. Rev.",
      volume         = "D46",
      year           = "1992",
      pages          = "539-549",
      doi            = "10.1103/PhysRevD.46.539",
      reportNumber   = "BA-92-11",
      SLACcitation   = "%%CITATION = PHRVA,D46,539;%%"
}

@article{Holman:1992us,
      author         = "Holman, Richard and Hsu, Stephen D. H. and Kephart,
                        Thomas W. and Kolb, Edward W. and Watkins, Richard and
                        Widrow, Lawrence M.",
      title          = "{Solutions to the strong CP problem in a world with
                        gravity}",
      journal        = "Phys. Lett.",
      volume         = "B282",
      year           = "1992",
      pages          = "132-136",
      doi            = "10.1016/0370-2693(92)90491-L",
      eprint         = "hep-ph/9203206",
      archivePrefix  = "arXiv",
      primaryClass   = "hep-ph",
      reportNumber   = "NSF-ITP-92-06, CMU-HEP92-05, FERMILAB-PUB-92-034-A,
                        HUTP-92-A011, VAND-TH-92-2",
      SLACcitation   = "%%CITATION = HEP-PH/9203206;%%"
}

@article{Kamionkowski:1992mf,
      author         = "Kamionkowski, Marc and March-Russell, John",
      title          = "{Planck scale physics and the Peccei-Quinn mechanism}",
      journal        = "Phys. Lett.",
      volume         = "B282",
      year           = "1992",
      pages          = "137-141",
      doi            = "10.1016/0370-2693(92)90492-M",
      eprint         = "hep-th/9202003",
      archivePrefix  = "arXiv",
      primaryClass   = "hep-th",
      reportNumber   = "IASSNS-HEP-92-9, PUPT-92-1309",
      SLACcitation   = "%%CITATION = HEP-TH/9202003;%%"
}

@article{Perl:2009zz,
      author         = "Perl, Martin L. and Lee, Eric R. and Loomba, Dinesh",
      title          = "{Searches for fractionally charged particles}",
      journal        = "Ann. Rev. Nucl. Part. Sci.",
      volume         = "59",
      year           = "2009",
      pages          = "47-65",
      doi            = "10.1146/annurev-nucl-121908-122035",
      reportNumber   = "SLAC-REPRINT-2012-022",
      SLACcitation   = "%%CITATION = ARNUA,59,47;%%"
}

@article{Perl:2004qc,
    author = "Perl, M. L. and Lee, E. R. and Loomba, D.",
    title = "{A Brief review of the search for isolatable fractional charge elementary particles}",
    reportNumber = "SLAC-PUB-11010",
    doi = "10.1142/S0217732304016019",
    journal = "Mod. Phys. Lett. A",
    volume = "19",
    pages = "2595--2610",
    year = "2004"
}

@article{Halyo:1999wq,
    author = "Halyo, V. and Kim, P. and Lee, E. R. and Lee, I. T. and Loomba, D. and Perl, M. L.",
    title = "{Search for free fractional electric charge elementary particles}",
    eprint = "hep-ex/9910064",
    archivePrefix = "arXiv",
    reportNumber = "SLAC-PUB-8283",
    doi = "10.1103/PhysRevLett.84.2576",
    journal = "Phys. Rev. Lett.",
    volume = "84",
    pages = "2576--2579",
    year = "2000"
}

@article{Kim:1979if,
      author         = "Kim, Jihn E.",
      title          = "{Weak Interaction Singlet and Strong CP Invariance}",
      journal        = "Phys. Rev. Lett.",
      volume         = "43",
      year           = "1979",
      pages          = "103",
      doi            = "10.1103/PhysRevLett.43.103",
      reportNumber   = "UPR-0120T",
      SLACcitation   = "%%CITATION = PRLTA,43,103;%%"
}

@article{Dine:1981rt,
      author         = "Dine, Michael and Fischler, Willy and Srednicki, Mark",
      title          = "{A Simple Solution to the Strong CP Problem with a
                        Harmless Axion}",
      journal        = "Phys. Lett.",
      volume         = "B104",
      year           = "1981",
      pages          = "199-202",
      doi            = "10.1016/0370-2693(81)90590-6",
      reportNumber   = "Print-81-0320 (IAS,PRINCETON)",
      SLACcitation   = "%%CITATION = PHLTA,B104,199;%%"
}

@article{Zhitnitsky:1980tq,
      author         = "Zhitnitsky, A. R.",
      title          = "{On Possible Suppression of the Axion Hadron
                        Interactions. (In Russian)}",
      journal        = "Sov. J. Nucl. Phys.",
      volume         = "31",
      year           = "1980",
      pages          = "260",
      note           = "[Yad. Fiz.31,497(1980)]",
      SLACcitation   = "%%CITATION = SJNCA,31,260;%%"
}

@article{Shifman:1979if,
      author         = "Shifman, Mikhail A. and Vainshtein, A. I. and Zakharov,
                        Valentin I.",
      title          = "{Can Confinement Ensure Natural CP Invariance of Strong
                        Interactions?}",
      journal        = "Nucl. Phys.",
      volume         = "B166",
      year           = "1980",
      pages          = "493",
      doi            = "10.1016/0550-3213(80)90209-6",
      reportNumber   = "ITEP-64-1979",
      SLACcitation   = "%%CITATION = NUPHA,B166,493;%%"
}

@article{Weinberg:1977ma,
      author         = "Weinberg, Steven",
      title          = "{A New Light Boson?}",
      journal        = "Phys. Rev. Lett.",
      volume         = "40",
      year           = "1978",
      pages          = "223-226",
      doi            = "10.1103/PhysRevLett.40.223",
      reportNumber   = "HUTP-77/A074",
      SLACcitation   = "%%CITATION = PRLTA,40,223;%%"
}

\end{document}